\documentclass[superscriptaddress,nofootinbib,aps,twocolumn,prc]{revtex4-1}

\usepackage{booktabs, ctable}
\usepackage{multirow}
\usepackage{amsmath,amssymb}
\usepackage{wasysym}
\usepackage{graphicx}
\usepackage{slashed}
\usepackage{url}
\usepackage{times}
\usepackage{xcolor}
\usepackage{bm}
\usepackage{color}
\usepackage{footnote}
\usepackage[symbol]{footmisc}
\usepackage{hyperref}

\usepackage[normalem]{ulem}  
\renewcommand\sout{\bgroup \color{red} \ULdepth=-.5ex \ULset}

\graphicspath{{./figs/}}

\begin{document}
\title{Wigner Phase-Space Densities of Nuclear Clusters and Hypernuclei}

\author{Jiaxing Zhao}
\affiliation{Institute for Theoretical Physics, Johann Wolfgang Goethe Universit\"{a}t, Frankfurt am Main, Germany}
\affiliation{Helmholtz Research Academy Hessen for FAIR (HFHF),GSI Helmholtz Center for Heavy Ion Research. Campus Frankfurt, 60438 Frankfurt, Germany}

\author{Joerg Aichelin}
\affiliation{SUBATECH UMR 6457 (IMT Atlantique,  Universit\'{e} de Nantes, IN2P3/CNRS), 4 Rue Alfred Kastler, F-44307 Nantes, France}

\author{Elena Bratkovskaya}
\affiliation{GSI Helmholtzzentrum f\"{u}r Schwerionenforschung GmbH, Planckstrasse 1, 64291 Darmstadt, Germany}
\affiliation{Institute for Theoretical Physics, Johann Wolfgang Goethe Universit\"{a}t, Frankfurt am Main, Germany}
\affiliation{Helmholtz Research Academy Hessen for FAIR (HFHF),GSI Helmholtz Center for Heavy Ion Research. Campus Frankfurt, 60438 Frankfurt, Germany}

\begin{abstract}
We solve the Schrödinger equation for few-body systems to obtain the wave function for light nuclear clusters and hypernuclei from d to $\rm ^5_{\Lambda\Lambda}He$ employing realistic nucleon-nucleon and nucleon-$\Lambda$ potentials. We project the solution to the hyperspherical harmonic basis states to obtain the corresponding density matrices and the Wigner densities. The experimental root mean square (rms) radii and binding energies of the different clusters are well reproduced. The Wigner densities obtained 
will allow to improve the present coalescence approaches to identify clusters, created in heavy-ion collisions.
\end{abstract}
\date{\today}

 \maketitle

\section{Introduction}
\label{sec.int}

Nuclear clusters (such as deuterons, tritons, and helium nuclei) and hypernuclei — nuclei containing one or more hyperons (e.g., $\Lambda$, $\Sigma$, or $\Xi$) — serve as essential tools for probing the strong interaction.
Understanding the formation mechanisms of these loosely bound objects is essential for exploring the  strongly interacting matter at the final stages of heavy-ion collisions, as well as for investigating the role of strangeness in dense nuclear environments. In particular, hypernuclei provide valuable insights into hyperon–nucleon (Y–N) and hyperon–hyperon (Y–Y) interactions, which are essential for understanding the structure of neutron stars and for addressing the so-called hyperon puzzle in nuclear astrophysics~\cite{Schaffner-Bielich:2008zws,Lattimer:2015nhk}.

At existing facilities such as RHIC and the LHC, experiments (e.g., STAR and ALICE) have measured cluster and hypernuclei yields across a wide range of beam energies~\cite{STAR:2023uxk,STAR:2021orx,ALICE:2024djx,ALICE:2022sco,ALICE:2024koa}. These results have deepened our understanding of baryon clustering, freeze-out dynamics, and the role of strangeness in the hadronic phase. Looking forward, upcoming experiments at FAIR (CBM), J-PARC, NICA, HIAF, will allow unprecedented access to the high baryon density and multi-strange regimes, enabling precise studies of nuclear cluster, hypernuclei, and potential new hadron states.

The production of clusters and hypernuclei in heavy-ion collisions is a complex process, involving the formation of few-body systems within a highly dynamic many-body environment.
To date, only phenomenological models exist for the formation of quantum objects such as clusters within semiclassical transport approaches. Three main mechanisms have been developed:
(1) {\it Coalescence}, where nucleons, which are closed in phase-space, form clusters at freeze-out, 
where the parameters are tuned to reproduce multiplicities;
(2) {\it Collision-induced formation}, where deuterons emerge via specific three-body reactions like $NNN \rightarrow dN$ or $NN\pi \rightarrow d\pi$, with production rates calculated from measured inverse cross sections assuming matrix elements depend only on $\sqrt{s}$~\cite{Oliinychenko:2018ugs,Oliinychenko:2020znl}; 
(3) {\it Potential-driven formation}, in which attractive baryonic potentials at low density dynamically lead to cluster formation during the expansion phase, enabling prediction of clusters of all sizes using only the interactions already present in the transport model~\cite{Aichelin:1991xy,Aichelin:2019tnk,Glassel:2021rod,Coci:2023daq,Kireyeu:2024hjo}. 
Experimentally, light nuclei and hypernuclei have been observed over a wide range of collision energies, from the AGS and SPS to RHIC and the LHC. Their yields and kinematic properties exhibit intriguing systematics that challenge conventional coalescence and thermal models~\cite{Andronic:2010qu,Steinheimer:2012tb}, motivating the development of more sophisticated dynamical approaches.

The coalescence model is a widely used theoretical framework to describe the formation of light nuclei in high-energy heavy-ion collisions. In this approach, clusters such as deuterons, tritons, helium-3, and alpha particles are formed when nucleons (protons and neutrons) are at freeze-out within a certain relative spatial (${\bf r}_0$) and relative momentum (${\bf p}_0$). This phenomenological model captures the essential features of cluster formation as a few-body process embedded in a many-body system, and it has been successful in explaining cluster yields across a wide range of collision energies~\cite{Sombun:2018yqh,Liu:2019nii,Kireyeu:2022qmv,Reichert:2025rnw}. For deuterons,  more sophisticated versions of the model use its Wigner phase-space density, enabling a quantum-mechanical treatment and a better sensitivity to the internal structure of the clusters. Most implementations adopt a Gaussian-form of the Wigner density, which assumes a 3-D isotropic harmonic oscillator interaction between nucleons. In this case, the width parameter of the Gaussian is either related to the root-mean-square (rms) radius of the cluster or treated as a free parameter tuned to experimental data~\cite{Scheibl:1998tk,Zhu:2015voa,Aichelin:1987rh,Sun:2020uoj,Wang:2025ify,Wang:2023rpd,Liu:2024ygk,Lin:2025dsm}. A realistic wavefunction or Wigner function has so far been employed only for deuteron production within the coalescence model~\cite{Bellini:2020cbj,Mahlein:2023fmx}.
Extending the coalescence framework to describe heavier clusters or hypernuclei presents significant challenges, primarily due to the difficulty to construct  a realistic Wigner density of the clusters but as well as due to the lack of a precise knowledge of their rms radii and the eventual contributions from excited states.

To date, numerous ab initio methods have been developed and successfully applied to solve nuclear many-body problems for both bound and scattering states. These include the hyperspherical harmonics expansion~\cite{Barnea:1999be,Marcucci:2019hml,Zhao:2020nwy}, variational methods~\cite{Kohn:1948col,demkov1963variational,Kamimura:1988zz}, and advanced numerical approaches based on Faddeev or Faddeev-Yakubovsky equations~\cite{Faddeev:1960su,glockle2012quantum}. For a review we refer to~\cite{Leidemann:2012hr,Ekstrom:2022yea,Papenbrock:2024vhr}. Recently, these approaches have also been extended to the study of hypernuclei~\cite{Wirth:2014apa,Wirth:2017lso,Le:2020zdu,Le:2022ikc,Le:2024rkd}. However, most of these studies focus primarily on the mass spectra and binding energies of nuclear clusters and hypernuclei. While important, this information is not sufficient for studying their production in heavy-ion collisions by coalescence models, as previously discussed, where the knowledge of the Wigner phase-space density of the clusters is required.

The goal of this study is to establish a bridge between ab initio calculations of nuclear clusters and hypernuclei and their production probabilities in heavy-ion collisions, with the Wigner density serving as the key connecting quantity. Within this framework, the coalescence probability for each cluster is derived directly from its wavefunction via Wigner transformation, thereby linking the production yield to the cluster’s intrinsic quantum properties without introducing additional free parameters. This method offers a rigorous basis for the study of cluster production by the coalescence mechanism at the end of a heavy-ion collision.

This paper is organized as follows: In Section~\ref{sec.wavef} we present our solution of the $N$-body Schr\"odinger equation. We discuss the potentials, which we employ, as well as the solution of the Schr\"odinger equation using the hyperspherical harmonic functions as basis states. In section~\ref{sec.wignerf} we present the Wigner density algorithm and display the Wigner densities for the clusters of interest. Finally, in Section~\ref{sec.summary} we draw our conclusion.

\section{Few-body bound states}
\label{sec.II}
From a theoretical perspective, solving the $N$-body bound-state problem remains a significant challenge, especially when realistic potentials with short-range repulsion and tensor components are included. In this section, we will present the framework for the solution of the $N$-body Schr\"odinger equation based on the hyperspherical harmonics formalism and phase-space distributions of nuclear clusters and hypernuclei. 

\subsection{$N$-body Schr\"odinger equation}
\label{sec.wavef}
The Hamiltonian of $N$-body system can be expressed as
\begin{eqnarray}
H=\sum_{i=1}^N{{\bf p}^2_i \over 2m_i}+\sum_{i<j} V_{ij}.
\end{eqnarray}
We assume that the interaction potential $V=\sum_{i<j}V_{ij}$ is the summation of the two-body interactions, and 
genuine three-body potentials are neglected. First, we introduce the Jacobi coordinates for a $N$-body system to transform the individual particle coordinates to the center-of-mass (CoM) coordinate ${\bf R}$ and relative coordinates ${\bm \chi}$ via,
\begin{eqnarray}
{\bf R}&=& {1\over M}\sum_{i=1}^Nm_i{\bf r}_i, \nonumber\\
{\bm \chi}_{N-j}&=&\sqrt{M_j m_{j+1}\over M_{j+1}\mu }\left( {\bf r}_{j+1}-{1\over M_j}\sum_{i=1}^j m_i {\bf r}_i\right),
\end{eqnarray}
where $M_j=\sum_{i=1}^j m_i$ and $j=1,...N-1$. ${\bf r}_i$ is the coordinate of nucleon $i$. $\mu$ is a parameter with the dimension of a mass. Its value does not affect the final results. 
The conjugate momenta transformation can be expressed as:
\begin{eqnarray}
{\bf P}&=&\sum_{i=1}^N {\bf p}_i, \nonumber\\
{\bf q}_{N-j}&=&\sqrt{M_j m_{j+1}\over M_{j+1}\mu }\left( {\mu \over m_{j+1}}{\bf p}_{j+1}-{\mu \over M_j}\sum_{i=1}^j {\bf p}_i\right),
\end{eqnarray}
where ${\bf P}$ and ${\bf q}_i$ are the total and relative momenta, respectively.
Under such a coordinate transformation, the total kinetic energy becomes
\begin{equation}
\sum_{i=1}^N {{\bf p}^2_i\over 2m_i} = \frac{{\bf P}^2}{2M}+\sum_{i=1}^{N-1}\frac{{\bf q}_i^2}{2\mu}.
\end{equation}
We take $\mu\equiv M$ in the following numerical calculations. Since the potential depends only on the relative coordinates ${\bm \chi}_i$, one can factorize the $N$-body motion into a CoM motion, which is described by a plane wave, and a relative motion,  $\Psi({\bf r}_1,...,{\bf r}_N)=\phi({\bf R})\Phi({\bm \chi}_1,...,{\bm \chi}_{N-1})$.The bound state properties relate only to the relative motion of the system, and we just need to deal with the $3(N-1)$-dimensional wave equation. We then transform the relative coordinates ${\bm \chi}_1$,...,${\bm \chi}_{N-1}$ into one hyper-radius, 
\begin{eqnarray}
\rho\equiv \sqrt{{\bm \chi}_1^2+...+{\bm \chi}_{N-1}^2}
\end{eqnarray}
and $3N-4$ hyper-angles $\Omega=\{\alpha_{N-1},...,\alpha_2, \theta_1,\phi_1,...,\theta_{N-1},\phi_{N-1}\}$.  where the angles $\alpha_i$ are defined by $\sin \alpha_{i}\equiv |{\bm \chi}_i|/\rho_i, \quad i=2,...,N-1.$, and are within the range $[0,\pi/2]$. $\rho_i=\sqrt{\sum_j^i  {\bm \chi}_j^2}$ and $\rho\equiv \rho_{N-1}$.
$\{\theta_i, \phi_i\}$ are the spherical coordinates corresponding to ${\bm \chi}_i$.

Now, the relative wavefunction satisfies the equation, 
\begin{eqnarray}
&&\left [ {1\over 2\mu}\left( -{1 \over \rho^{3N-4}}{d\over d\rho}\rho^{3N-4}{d \over d\rho}  + {\hat {\bf K}_N^2\over \rho^2}\right) + V(\rho, \Omega) \right ]\Phi ,\nonumber\\
&&= E_r \Phi
\label{eq.relamom}
\end{eqnarray}
 where $\hat {\bf K}_N$ is the hyperangular momentum operator and $E_r$  
 $E_r$ is, if the potential approaches zero at infinity, equal in magnitude but opposite in sign to the binding energy, which is conventionally defined as positive.
The potential $V(\rho, \Omega)$ depends not only on the hyper-radius but also on the $3N-4$ hyper-angles. In this case, the Schr\"odinger equation~\eqref{eq.relamom} cannot be further factorized into a radial part and an angular part. Instead, one can expand the wave function in terms of the hyperspherical harmonic (HH) functions ${\mathcal Y}_\kappa(\Omega)$, which are the eigenstates of the hyper-angular momentum operator $\widehat {\bf K}_N^2$ and form a complete basis,
\begin{equation}
\widehat {\bf K}_N^2{\mathcal Y}(\Omega)=K_N(K_N+3N-5){\mathcal Y}(\Omega).
\end{equation}
$K_N$ is the grand angular quantum number, which is the hypermomentum in hyperradial basis. Properties of the hyperspherical harmonic functions ${\mathcal Y}_\kappa(\Omega)$  can be found in Refs.~\cite{Barnea:1999be,Barnea:2006sd,Zhao:2023qww,Zhao:2017gpq}.
There are eight operators that commute with the kinetic energy term in the Hamiltonian and with each other. In addition to $K_N$, the other conserved quantum numbers are the total orbital angular momentum ($L$), the total magnetic quantum number ($M$). The angular momentum, corresponding to each Jacobi coordinate ($l_1$,...,$l_{N-1}$), is the orbital angular momentum quantum number of each subsystem of particle $i$ and $j$. The relative radial quantum number $n_2$,...,$n_{N-1}$. These quantum numbers satisfy,
\begin{eqnarray}
&&K_j=2n_j+K_{j-1}+l_j, \quad \quad   (j=1,...,N) ,\nonumber\\
&&L=\sum_i^{N-1}l_i,\quad M=\sum_i^{N-1}m_i,
\end{eqnarray}
where $K_0=0$ and $n_1=0$ by definition. We can introduce the shorthand notation $\kappa$, which stands for all the quantum numbers mentioned above. 
The HH function can be expressed as, 
\begin{eqnarray}
{\mathcal Y}_{\kappa}(\Omega) &=& \Big[ \sum_{m_1,...,m_n} \langle l_1m_1l_2m_2|L_2M_2\rangle \langle L_2M_2l_3m_3|L_3M_3 \rangle \nonumber\\
&\times& ... \langle L_{n-1}M_{n-1}l_nm_n|L_nM_n \rangle  \prod_{j=1}^n Y_{l_j,m_j}(\theta_j,\phi_j) \Big] \nonumber\\
& \times & \Big[ \prod_{j=2}^n {\mathcal N}_j (\sin \alpha_j)^{l_j} (\cos \alpha_j)^{K_{j-1}} \nonumber\\
& \times &P_{n_j}^{l_j+1/2, K_{j-1}+(3j-5)/2}(\cos 2 \alpha_j) \Big],
\end{eqnarray}
where $n=N-1$. $\hat L_i=\hat L_{i-1}+\hat l_i$ and $\hat M_i=\hat M_{i-1}+\hat m_i$. $\langle l_1m_1l_2m_2|L_2M_2\rangle$ are the Clebsch-Gordan coefficients. $Y_{l_j,m_j}$ is the ordinary spherical harmonic function. $P_{n}^{a,b}$ are  the Jacobi polynomials. ${\mathcal N}_j$ are normalization constants given by
\begin{eqnarray}
{\mathcal N}_j =\sqrt{ (2K_j+3j-2)n_j! \Gamma(n_j+K_{j-1}+l_j +{3j-2\over 2}) \over \Gamma(n_j+l_j+{3\over 2}) \Gamma(n_j+K_{j-1}+{3j-3\over 2})}.
\end{eqnarray}

Then, the relative wavefunction can be expressed as, 
 \begin{eqnarray}
 \Phi(\rho, \Omega)=\sum_\kappa R_{\kappa}(\rho){\mathcal Y}_\kappa(\Omega).
\end{eqnarray}
Introducing this in the  equation for the relative motion~\eqref{eq.relamom}, one can obtain a set of coupled differential equations,
\begin{eqnarray} 
&&\Bigg[{1\over 2\mu}\left({d^2 \over d\rho^2}+{3N-4\over \rho}{d \over d\rho} - {K_N(K_N+3N-5)\over \rho^2}\right) \nonumber\\
&&+E_r \Bigg]R_{\kappa}=\sum_{\kappa'}V_{\kappa \kappa'}R_{\kappa'}.
\end{eqnarray}
If we further simplify the radial wavefunction as $R_\kappa(\rho)=u_\kappa(\rho)/\rho^{(3N-4)/2}$. we obtain
\begin{eqnarray} 
&&\Bigg[{1\over 2\mu}\left({d^2 \over d\rho^2}-{\mathcal {K}\over \rho^2} - {K(K+3N-5)\over \rho^2}\right)+E_r \Bigg]u_{\kappa}\nonumber\\&&=\sum_{\kappa'}V_{\kappa \kappa'}u_{\kappa'},
\label{eq.coupn}
\end{eqnarray}
where $\mathcal {K}\equiv 3(3N^2-10N+8)/4$. $V_{\kappa \kappa'}$ is the potential matrix, 
\begin{eqnarray}
V_{\kappa \kappa'}&=&\int {\mathcal Y}_\kappa^*(\Omega)V(\rho, \Omega){\mathcal Y}_{\kappa'}(\Omega)d\Omega \nonumber\\
	&=&\sum_{i<j}\int V^{(ij)}(|\mathbf{r}_i-\mathbf{r}_j|){\mathcal Y}_\kappa^*(\Omega){\mathcal Y}_{\kappa'}(\Omega)\mathrm{d}\Omega,
\end{eqnarray}

with the volume element for $N$-body case 

\begin{eqnarray}
d\Omega= \Bigg[\prod_{j=1}^{N-1}\sin \theta_j d\theta_jd\phi_j\Bigg] \prod_{j=2}^{N-1} (\sin \alpha_j)^2 (\cos \alpha_j)^{3j-4}d\alpha_j.
\end{eqnarray}
For the three and four body case, the potential can be calculated by transforming the HH basis to another basis considering the Raynal-Revai coefficients~\cite{Raynal:1970ah,Zhao:2023qww}.
This coupled differential equation~\eqref{eq.coupn} can be solved numerically via e.g. the inverse power method~\cite{Zhao:2024atb}.

The radial probability of each component is defined as,
\begin{eqnarray}
P_\kappa(\rho)\equiv |R_\kappa(\rho)|^2\rho^{3N-4}.
\label{eq.defPrho}
\end{eqnarray}
The normalization condition is fulfilled, $\int \sum_\kappa P_\kappa(\rho)d\rho=1$.
The root-mean-squared radius (rms) of the $N$-body cluster is defined as
\begin{eqnarray}
r_{\rm rms}^2\equiv \langle \rho^2\rangle=\int \sum_\kappa |R_\kappa(\rho)|^2\rho^{3N-2} d\rho.
\end{eqnarray}
From the definition of the hyperradius $\rho$, we have,
\begin{eqnarray}
 \langle \rho^2\rangle&=& \langle \sum_{j=1}^{N-1}{\bm \chi}_j^2\rangle \nonumber\\
&=&{1\over \mu M }\sum_{i<j} m_im_j  \langle {\bf r}_{ij}^2\rangle \nonumber\\
&=&{1\over \mu}\sum_{i=1}^N m_i  \langle ({\bf r}_i-{\bf R})^2\rangle.
\label{eq.rms}
\end{eqnarray}
If the particles have the same mass, the root-mean-square radius represents the geometric radius $r_{\rm geo}^2=1/N\sum_{i=1}^N \langle ({\bf r}_i-{\bf R})^2\rangle$ of the cluster.
The definition of the cluster mass and binding energy are 
\begin{eqnarray}
\rm M&\equiv& \sum_i^N m_i+E_r\nonumber\\
\rm B.E.&\equiv& -E_r.
\label{eq.defM}
\end{eqnarray}
The mass of proton, neutron, and hyperon $\Lambda$ are taken as 0.938 GeV, 0.939 GeV, and 1.116 GeV~\cite{ParticleDataGroup:2024cfk}, respectively.  

\begin{figure}[!htb]
\includegraphics[width=0.45\textwidth]{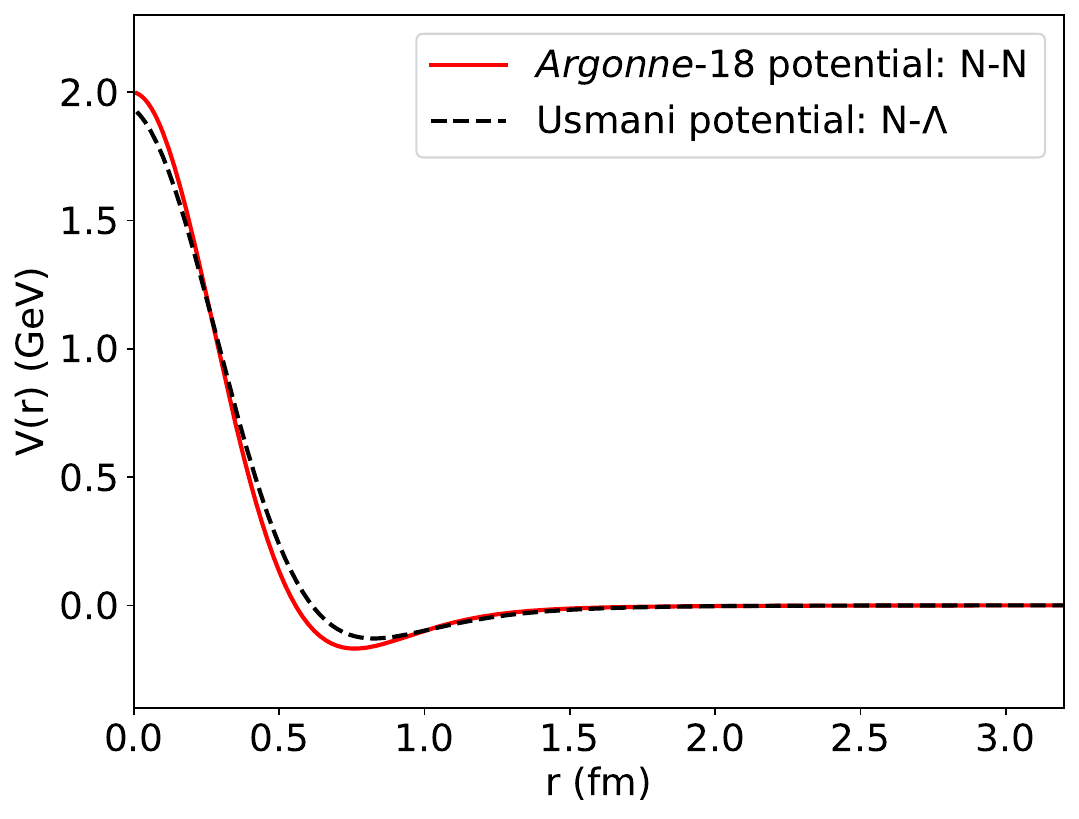}
\caption{The interaction potential between nucleons described by the \textit{Argonne}-18 potential~\cite{Wiringa:1994wb}, and between a nucleon and a $\Lambda$ described by the Usmani potential~\cite{Bodmer:1984gc}. The parameters of these forms are determined by fitting the N-N elastic scattering data.}
\label{fig.pot}
\end{figure}

\emph{Two-body.--}The deuteron is the simplest bound nuclear system, consisting of a proton and a neutron. Its mass is approximately 1875.613 MeV, slightly less than the combined mass of a free proton and neutron due to the small binding energy 2.224MeV. The two-body Schr\"odinger equation has been successful used to describe its mass and root-mean-radius~\cite{Lamia:2012zz,Mahlein:2023fmx}. The radial equation of the two-body Schr\"odinger can be written as,
\begin{eqnarray} 
\Bigg[{1\over 2\mu}\left({d^2 \over d\rho^2}- {L(L+1)\over \rho^2}\right)+V_{12} \Bigg]u(\rho) =E_ru(\rho).
\end{eqnarray}
$V_{12}$ is the interaction between nucleons, which can be described in the Yukawa theory~\cite{Yukawa:1935xg} and has been  extended to the one-boson-exchange (OBE) model~\cite{Bryan:1964zzb}. 
There are numerous OBE-based/extended phenomenological nucleon-nucleon (N-N) potentials, including the Paris potential~\cite{Lacombe:1980dr}, the \textit{Argonne}-18 potential~\cite{Wiringa:1994wb}, the Reid Soft-Core potential~\cite{Reid:1968sq}, the Nijmegen potentials~\cite{Stoks:1994wp,Nagels:2014qqa}. The parameters of these potentials are determined by fitting the N-N elastic scattering data. In our calculation we employ for the interaction between nucleons the \textit{Argonne}-18 potential~\cite{Wiringa:1994wb} and, for the hyperon-nucleon (Y-N)  interaction, the potential from Usmani~\cite{Bodmer:1984gc}. Both are shown in Fig.~\ref{fig.pot}. Both are very similar due to the flavor blindness of QCD (apart from the masses). We see a soft repulsive core at short distances, which can be understand on the level of Quantum chromo dynamics by a combined effect of quark Pauli blocking, one-gluon exchange and  hidden color configurations~\cite{10.1143/PTPS.42.39}. The potential at middle and long range is attractive due to the exchange of $\pi$, and $\sigma$ mesons. 
In this paper, we neglect the spin-related potential, which only give a fine structure of the energy level, and take a spin-averaged potential to solving the Schr\"odinger equation.  

\begin{figure*}[!htb]
\includegraphics[width=1\textwidth]{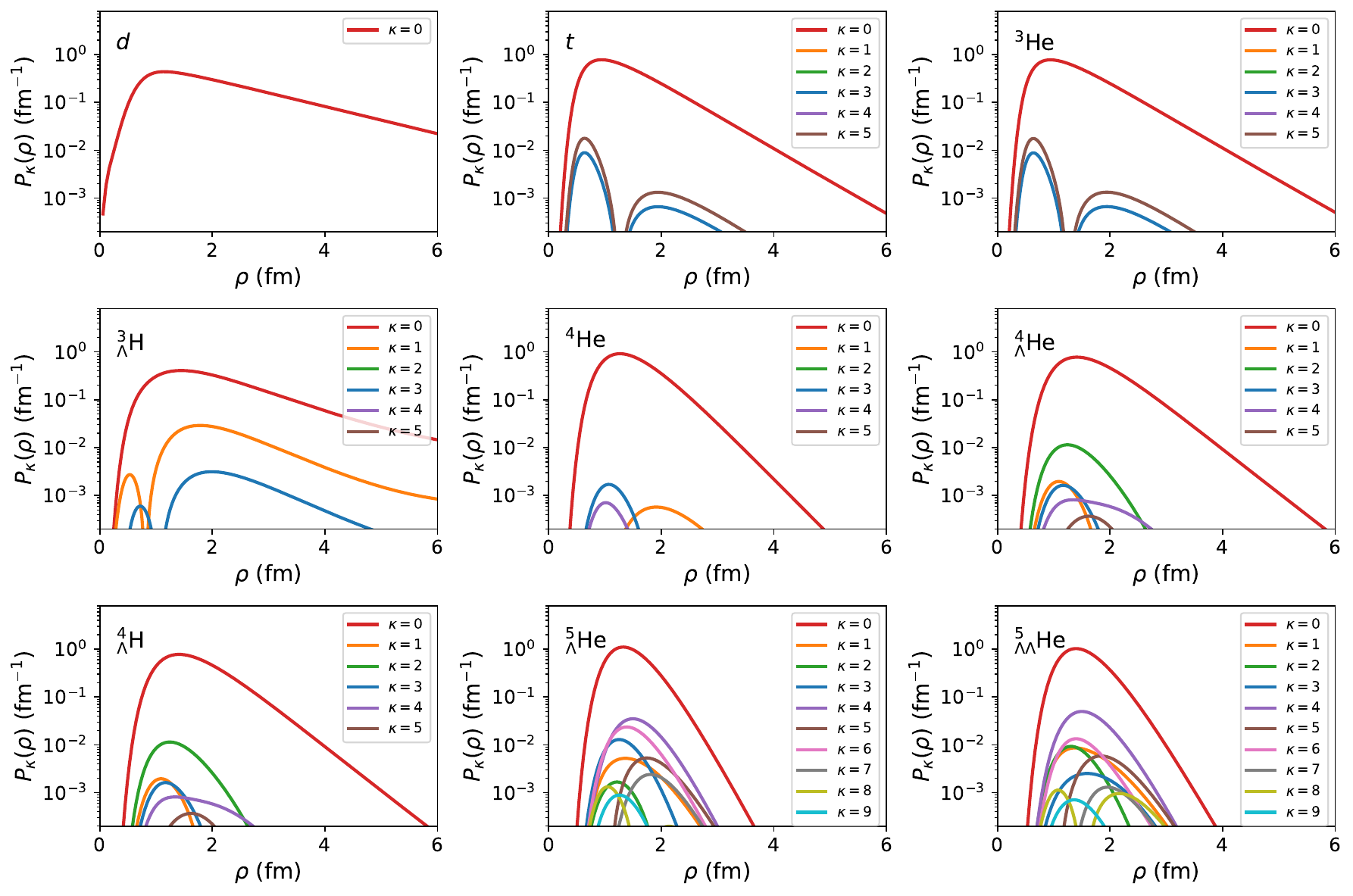}
\caption{The radial distribution $P_\kappa(\rho)$, as defined in Eq.~\eqref{eq.defPrho}, of $S$-wave $d$, $t$, $^3\rm He$, $^3_\Lambda \rm H$, $^4\rm He$, $^4_\Lambda \rm He$, $^4_\Lambda \rm H$, $^5_\Lambda \rm He$, and $^5_{\Lambda\Lambda} \rm He$. Each component with a different $ \kappa$ is displayed by different color.}
\label{fig.wfall}
\end{figure*}

By solving the radial equation, we obtain the wavefunction and the mass of a deuteron. The radial probability of deuteron is shown in the Fig.~\ref{fig.wfall}, top right. The obtained mass and the rms radius of the deuteron is shown in Table~\ref{tab2}. We can see that the masses defined in Eq.~\eqref{eq.defM} our results are very close to the experimental data and other model calculations~\cite{McGee:1967zza,Lacombe:1980dr,Bellini:2020cbj,Mahlein:2023fmx,Coci:2023daq}. 

\emph{Three-body.--}Triton, denoted as $t$ or $\rm ^3H$, is the nucleus of the hydrogen isotope tritium, consisting of one proton and two neutrons, which was first produced and identified in 1934 on nuclear fusion reactions involving deuterium. Its mass is 2808.921 MeV and the binding energy is 8.482 MeV,  making it more tightly bound than the deuteron. 
Helium-3, denoted as $\rm ^3He$, is a light, stable isotope of helium consisting of two protons and one neutron, which is detected through mass spectrometry. Its mass is close to that of the triton, 2808.391 MeV. The binding energy is 7.718 MeV. 
Another three-body cluster is the Hypertriton, denoted as $\rm _\Lambda^3H$, the lightest known hypernucleus, consisting of one proton, one neutron, and one $\Lambda$ hyperon. It was first discovered in 1952. As a weakly bound three-body system containing a strange quark, the hypertriton serves as a unique laboratory for exploring hyperon-nucleon (Y-N) interactions and strangeness in nuclear matter. The mass of $\rm _\Lambda^3H$ is 2991.16 MeV. It has a small binding energy, $\sim 2.36$ MeV~\cite{E864:2002xhb}.

For a system of three nucleons with realistic interactions there exists no analytical solution of the Schrödinger equation anymore and we have to solved it numerically, employing Eq.~\eqref{eq.coupn}. In actual calculations, the HH basis needs to be truncated, what makes the solution approximative. How to choose the truncation and how many basis states are needed depends on the symmetry of the system. 
It is the advantage of the hyperspherical expansion that we need only a very limited number number of basis states to obtain results, which are very close to the experimental values.  This can also inferred from Fig.~\ref{fig.wfall}.

For the $S$-wave, we can choose the HH basis with total angular momentum $L=0$. In this case, we take all possible basis states with $L=0$ and $K\leq 4$. There are 6 in total. For the $P$-wave we take $L=1$ and $K\leq 3$. The wave functions, masses and rms radii, which we obtain, are shown in Fig.~\ref{fig.wfall} and Table~\ref{tab2}. We can see that the masses, which we obtain with this truncation, are very close to the experimental values. $P$-wave excited states are not founded in our study.

\emph{Four-body.--} Helium-4, denoted as $\rm ^4He$, is one of the most stable and abundant light nuclei in the universe, consisting of two protons and two neutrons. It was first detected in 1868 in the solar spectrum. The mass of $\rm ^4He$ is 3727.38 MeV with a very large binding energy, 28.3 MeV, indicating that  it is extremely stable and tightly bound. The other two interesting four body bound systems are Hyperhydrogen-4 ($\rm ^4_\Lambda H$) and Hyperhelium-4 ($\rm ^4_\Lambda He$). They are isospin mirror partners. Their masses are 3922.8 MeV and 3922.8 MeV, with binding energies of 2.04 and 2.39 MeV, respectively. Due their small production cross sections
these hypernuclei are rarely produced and therefore their masses are not precisely known so far.  

We solved the coupled wave equation Eq.~\eqref{eq.coupn} numerically for these four-body bound states. For the $S$-wave, we can choose the HH basis with total angular momentum $L=0$. In this case, we choose all possible basis states with $L=0$ and $K\leq 4$. There are 8 in total. For the $P$-wave we take $L=1$ and $K\leq 2$. The obtained wavefunctions and masses are shown in Fig.~\ref{fig.wfall} and Table~\ref{tab2}. We can see that also their masses are very close to the experimental values. No any $P$-wave excited states in our calculation.

\emph{Five-body.--}Hyperhelium-5, $\rm ^5_\Lambda He$, is a light hypernucleus consisting of two protons, two neutrons, and one $\Lambda$ (lambda) hyperon. Its mass is 4731.5 MeV with the binding energy is 3.12 MeV. Another five body hypernucleus is double-$\Lambda$ hypernucleus, $\rm ^5_{\Lambda\Lambda} He$, which is predicted by many theoretical models~\cite{Nemura:1999qp} but still unconfirmed experimentally. To solve the coupled wave equation Eq.~\eqref{eq.coupn}, we choose all possible basis states with $L=0$ and $K\leq 2$, 10 in total, for the $S$-wave, the ground state. For the $P$-wave we take $L=1$ and $K\leq 2$ but no $P$-wave excited states are not founded.

\begin{table*}[!hptb]
\renewcommand\arraystretch{2.0}
\setlength{\tabcolsep}{4.0mm}
\begin{tabular}{c|c|ccc|ccc|cc}
	\toprule[1pt]\toprule[1pt] 
        Cluster &$d$ & $t$ & $\rm ^3He$ & $\rm ^3_\Lambda H$ & $\rm ^4 He$ & $\rm ^4_\Lambda He$ & $\rm ^4_\Lambda H$ & $\rm ^5_\Lambda He$ & $\rm ^5_{\Lambda \Lambda}He$ \\	
	\toprule[1pt] 
	Constitutes & pn & pnn & ppn & $\rm pn\Lambda$ & ppnn & $\rm ppn\Lambda$ & $\rm pnn\Lambda$ & $\rm ppnn\Lambda$ & $\rm ppn\Lambda\Lambda$\\
        $J^P$ & $1^+$ & ${1\over 2}^+$ & ${1\over 2}^+$ & ${1\over 2}^+$ & $0^+$ & $0^+$ & $0^+$ & ${1\over 2}^+$ & ${1\over 2}^+$\\
        \toprule[1pt] 
	$\rm M_{exp.}$(GeV) & 1.875 & 2.809  & 2.808 & 2.991  & 3.727 & 3.923 & 3.923 & 4.731 & -\\
        \toprule[1pt] 
        $\rm M_{theo.}$(GeV) & 1.873 & 2.813 & 2.812 & 2.993 & 3.746 & 3.927 & 3.929 & 4.847 & 5.028\\
        rms(fm)  & 2.790 & 1.561 & 1.567 & 4.332 & 1.590 & 1.810 & 1.809 & 1.509 & 1.595 \\        
	\bottomrule[1pt]\bottomrule[1pt]
\end{tabular}
\caption{The experimental masses $\rm M_{exp.}$ and the calculated masses $\rm M_{theo.}$, and root-mean-square radii $\rm rms$ of different nuclear clusters and hypernuclei.}
\label{tab2}
\end{table*}

Because $\rm ^5Li$ clusters have not been measured experimentally at higher beam energies we refrain from including them in our analysis.

\subsection{$N$-body Wigner function}
\label{sec.wignerf}
The Wigner function is a quasi-probability distribution that provides a phase-space representation of a quantum state. As the classical phase space  distribution it combines position and momentum information into a single-valued function, making it particularly useful for visualizing quantum states and for semiclassical approximations.

The corresponding $N$-body Wigner function can be obtained from the $N$-body density matrix via the Wigner transformation, 
\begin{eqnarray}
&&W_N({\bf r}_1,...,{\bf r}_N; {\bf p}_1,...,{\bf p}_N) = \int d^{3} \mathbf{s}_1...d^{3} \mathbf{s}_N \nonumber\\
&&\times e^{ -i \sum_{k=1}^{N} {\bf p}_k \cdot \mathbf{s}_k } \, \Psi^*\left( {\bf r}_1 + \frac{\mathbf{s}_1}{2}, \dots, {\bf r}_N + \frac{\mathbf{s}_N}{2} \right) \nonumber\\
&&\times \Psi\left( {\bf r}_1 - \frac{\mathbf{s}_1}{2}, \dots, {\bf r}_N - \frac{\mathbf{s}_N}{2} \right).
\end{eqnarray}
It satisfies the normalization condition, 
\begin{eqnarray}
\prod_{i=1}^N \int  {d^3{\bf p}_i d^3 {\bf r}_i \over (2\pi)^3} W_N({\bf r}_1,...,{\bf r}_N; {\bf p}_1,...,{\bf p}_N) = 1.
\end{eqnarray}
The CoM motion is a plane wave and can be removed. Then the $N$-body Wigner function can be expressed with the relative coordinates as
\begin{eqnarray}
&&W_N({\bm \chi}_1,...,{\bm \chi}_{N-1}; {\bf q}_1,...,{\bf q}_{N-1}) =\int d^{3} \mathbf{s}_1...d^{3} \mathbf{s}_{N-1}   \nonumber\\
&&\times  e^{ -i \sum_{k=1}^{N-1} \mathbf{q}_k \cdot \mathbf{s}_k } \Phi^*\left({\bm \chi}_1 + \frac{\mathbf{s}_1}{2}, \dots, {\bm \chi}_N + \frac{\mathbf{s}_N}{2} \right) \nonumber\\
&&\times \Phi\left({\bm \chi}_1 - \frac{\mathbf{s}_1}{2}, \dots, {\bm \chi}_N - \frac{\mathbf{s}_N}{2} \right).
\end{eqnarray}
This Wigner function provides a quasi-probability distribution in the $6N$-dimensional phase space, combining positions and momenta. The solution of the $N$-body relative wavefunction $\Phi$ has been presented in last section.

Assuming a Gaussian form of the wave functions the Wigner function can be computed analytically and takes a Gaussian form in both coordinate and momentum space. This form is widely used in coalescence models for nuclear cluster and hypernuclei production in heavy-ion collisions. A Gaussian form of the wave function is obtained if the interaction between two particles is a 3-D isotropic harmonic oscillator potential.

\begin{figure*}[!htb]
\includegraphics[width=1\textwidth]{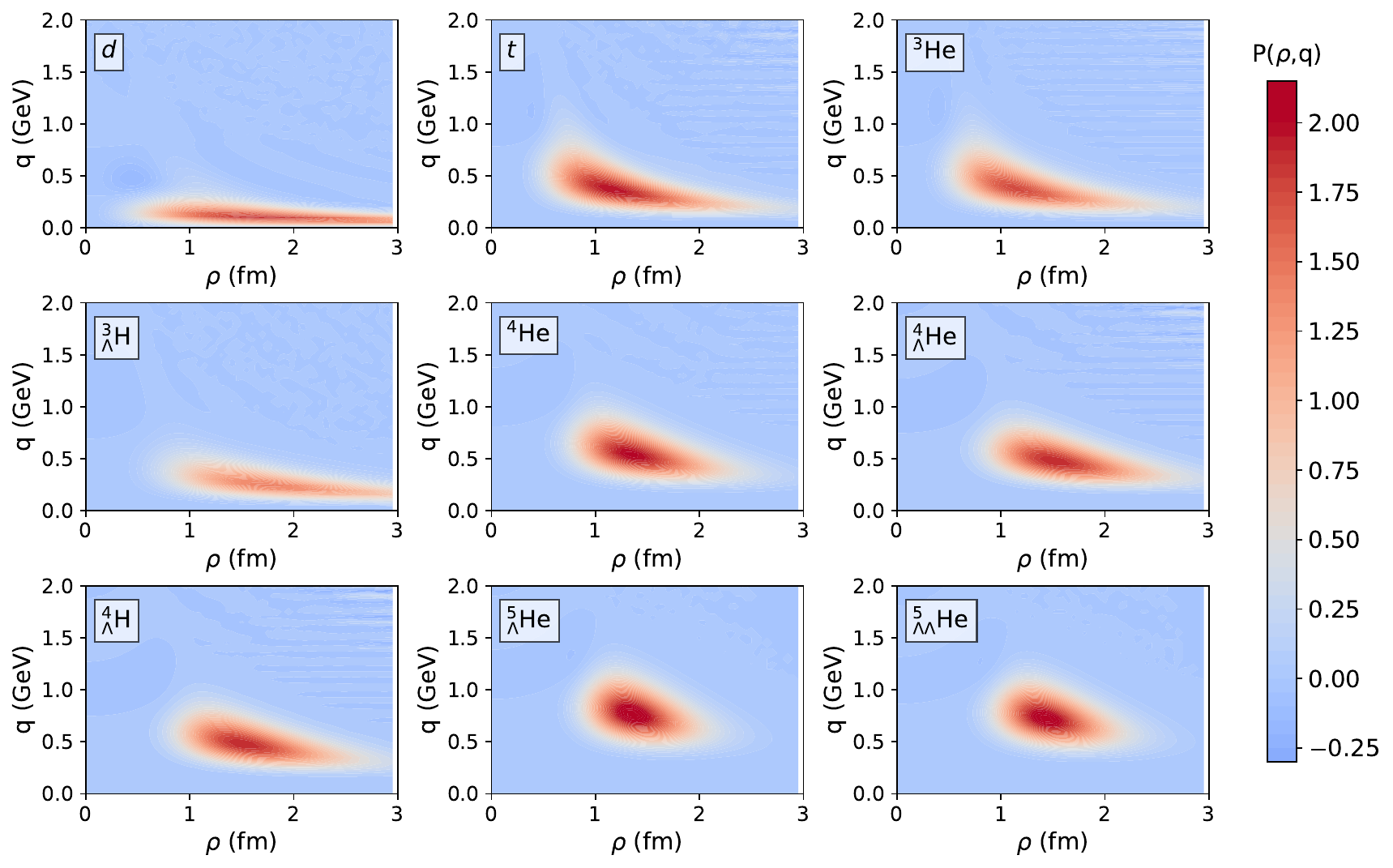}
\caption{Wigner density $P_N(\rho,q)$, as defined in Eq.~\eqref{eq.wignerProb}, of $S$-wave $d$ ($N=2$), $t$, $^3\rm He$, $^3_\Lambda \rm H$ ($N=3$), $^4\rm He$, $^4_\Lambda \rm H$, $^4_\Lambda \rm He$ ($N=4$), $^5_\Lambda \rm He$, and $^5_{\Lambda\Lambda} \rm He$ ($N=5$), where $\rho$ and $q$ are the hyperradius and its conjugate relative momentum, respectively.}
\label{fig.wignerall}
\end{figure*}
We will study this in detail. Let us first discuss the two-body case assuming that the potential between the heavy quark pairs is 3-D isotropic harmonic oscillator potential, $V(r)=1/(2\mu \sigma^4)r^2$. Then the wavefunction can be obtained in an analytically form (here we have $\sqrt{\mu\omega}=1/\sigma$). The wavefunction of a 3-D isotropic harmonic oscillator is given by $\psi_{nlm}(r,\theta,\phi)=R_{nl}(r)Y_{l,m}(\theta,\phi)$, where $Y_{l,m}$ is the spherical harmonics. 
The radial part can be expressed as,
\begin{eqnarray}
R_{nl}(r)&=&\left [{2 (n!) \over \sigma^3 \Gamma(n+l+3/2)} \right]^{1\over 2}\left({r\over \sigma} \right)^l e^{-{r^2\over 2\sigma^2}}\nonumber\\
&\times&L_n^{l+1/2}\left({r^2\over \sigma^2}\right),
\label{eq.3dwf}
\end{eqnarray}
where $L_n^{l+1/2}$ are Laguerre polynomials. The energy is $E_r=(2n+l+3/2)\omega=(2n+l+3/2)/(\mu\sigma^2)$.
For the ground state, the Wigner function is
\begin{eqnarray}
W_2^{\rm 1S}({\bf r},{\bf q})=8e^{-{{\bf r}^2\over \sigma^2}-{\bf q}^2\sigma^2}.
\end{eqnarray}
For the excited states, we refer to Ref.~\cite{Zhao:2023dvk}. The width $\sigma$ is related to the root-mean-square radius via,
\begin{eqnarray}
\langle r^2\rangle={3\over 2}\sigma^2, \quad   {\rm for ~ \ 1S}
\end{eqnarray}
For the $N$-body case we know from the Eq.~\eqref{eq.rms}
\begin{eqnarray}
\sum_{j=1}^{N-1}{\bm \chi}_j^2 ={1\over \mu \sum_{i=1}^N m_i }(\sum_{i<j} m_im_j {\bf r}_{ij}^2).
\end{eqnarray}
The $N$-body Schr\"odinger equation has an analytical solution if, similar to two-body case, the interaction potential has the form
\begin{eqnarray}
V({\bf r}_{ij})={1\over 2\mu \sigma_{ij}^4}{m_im_j\over \mu\sum_{i=1}^N m_i } {\bf r}_{ij}^2.
\label{eq.vij}
\end{eqnarray}
where the width $\sigma_{ij}$ is related to the two particle system ij and all widths should be equal. Then the relative motion can be expressed as,
\begin{eqnarray}
\sum_{i=1}^{N-1}\left(\frac{{\bf q}_i^2}{2\mu}+{1\over 2\mu \sigma^4}{\bm \chi}_i^2 \right)\Phi=E_r\Phi.
\end{eqnarray}
In this case, we can separate the relative wavefunction as a product of $N-1$ wavefunctions of a 3-D isotropic harmonic oscillator, 
 \begin{eqnarray}
\Phi=\prod_{i=1}^{N-1}\psi_{n_{{\bm \chi}_i},l_{{\bm \chi}_i},m_{{\bm \chi}_i}}({\bm \chi}_i)
\end{eqnarray}
$\psi_{n_i,l_i,m_i}(r,\theta,\phi)=R_{n_i,l_i}(r)Y_{l_i,m_i}(\theta_i,\phi_i)$. $R_{n_i,l_i}(r)$ is given by Eq.~\eqref{eq.3dwf}. 
The energy eigen value for a $N$-body system with this potential is 
 \begin{eqnarray}
E_r={1\over \mu\sigma^2}\sum_{i=1}^{N-1}(2n_{{\bm \chi}_i}+l_i+{3\over 2}).
\end{eqnarray}
With this analytical $N$-body wavefunction, the total Wigner function can easily be constructed as a product of two-body Wigner function $W_2$,
\begin{eqnarray}
W_N^{\rm 1S}({\bm \chi}_1,...,{\bm \chi}_{N-1}; {\bf q}_1,...,{\bf q}_{N-1})=\prod_{i=1}^{N-1}W_2^{\rm 1S}({\bm \chi}_i,{\bf q}_i).
\end{eqnarray}
Moreover, the total Wigner function of the ground state can be simplified as, 
\begin{eqnarray}
W_N^{\rm 1S}(\rho,q)=8^{N-1} e^{-{\rho^2\over \sigma^2}-q^2\sigma^2}.
\end{eqnarray}
For this analytically wavefunction the width $\sigma$ of the 1S state is related to the rms via
\begin{eqnarray}
\langle \rho^2 \rangle={3(N-1)\over 2}\sigma^2, 
\end{eqnarray}
Previous works on cluster production~\cite{Chen:2003ava,Sun:2015jta,Liu:2024ilw,Wang:2025ify} have approximated the three-body Wigner function by 
\begin{eqnarray}
W_3^{\rm 1S}({\bm \chi}_1,{\bm \chi}_{2}; {\bf q}_1,{\bf q}_2)=W_2^{\rm 1S}({\bm \chi}_1,{\bf q}_1)W_2^{\rm 1S}({\bm \chi}_2,{\bf q}_2).
\label{eq:twobytwo}
\end{eqnarray}
The quality of such an approximation is difficult to judge without knowing the solution of the Schrödinger equation. 
As we have seen,  in our approach the coupled equations can only be factorized if the interaction takes the specific form of Eq.~\eqref{eq.vij}, which implies that the interaction potential differs for pairs with different masses. In the general case, however, the $N$-body Schr\"odinger equation cannot be reduced to $N-1$ independent two-body problems.

In most of the cases, the higher-order components of the wavefunction are very small, and the lowest component with $\kappa=0$ dominates as shown in Fig.~\ref{fig.wfall}. This means that the total wavefunction can well be approximated by $\Phi(\rho,\Omega)\approx R_0(\rho)\mathcal Y_0(\Omega)$,
where $\mathcal Y_{0}$ is the ground state ($\kappa =0$) hyperspherical harmonic (HH) function, which means all quantum numbers equal to zero and is independent of the hyperangles $\Omega$. 

For a $N$-body system, the ground HH function can be expressed as
\begin{eqnarray}
\mathcal Y_{0}^N=\sqrt{\Gamma[3(N-1)/2]\over 2 \pi^{3(N-1)/2}}.
\end{eqnarray}
Then the Wigner transformation is simplified
\begin{eqnarray}
W_N({\bm \rho},{\bf q})&=&\int d^{3(N-1)}{\bf s} e^{-i{\bf q}\cdot {\bf s}}R_0^*(|{\bm \rho}+{{\bf s}\over2}|)({\mathcal Y}_{0}^{N})^* \nonumber\\ 
&\times &R_0(|{\bm \rho}-{{\bf s}\over2}|){\mathcal Y}_{0}^N.
\end{eqnarray}
Both $\bf q$ and $\bf s$ are vectors in the $3(N-1)$-dimensional space.
Because the wavefunction depends only on the hyperradius with rotational symmetry, we can simplify the integration 
\begin{eqnarray}
d^{3(N-1)}{\bf s}=s^{3N-4}ds (\sin\theta)^{3N-5}d\theta d\Omega_{3N-4}.
\end{eqnarray}
We define
\begin{eqnarray}
\mathcal S_{d}\equiv \int d\Omega_{d}={2\pi^{d/2}\over \Gamma[d/2]}.
\end{eqnarray}
Consequently:
\begin{eqnarray}
W_N({\bm \rho},{\bm q})&=&{\mathcal S_{3N-4}\over \mathcal S_{3N-3}}\int s^{3N-4}ds (\sin \theta_{qs})^{3N-5}d\theta_{qs} \nonumber\\ 
&\times &e^{-iqs\cos \theta_{qs}}R_0^*(|{\bm \rho}+{{\bf s}\over2}|)R_0(|{\bm \rho}-{{\bf s}\over2}|),
\end{eqnarray}
where $\theta_{qs}$ is the angle between vector $\bf q$ and $\bf s$. If integrated over all angles, we get an angle-independent Wigner function,
\begin{eqnarray}
W_N(\rho,q)&=&{\mathcal S_{3N-4}\over \mathcal S_{3N-3}}\int(\sin \theta_{\rho s})^{3N-5}d\theta_{\rho s} \nonumber\\ 
&\times &\int s^{3N-4}ds (\sin \theta_{qs})^{3N-5}d\theta_{qs} \nonumber\\ 
&\times &e^{-iqs\cos \theta_{qs}}R_0^*(\rho_+)R_0(\rho_-),
\label{eq.wignNN}
\end{eqnarray}
where $\rho_\pm=\sqrt{\rho^2+s^2/4\pm \rho s\cos \theta_{\rho s}}$ and $\theta_{\rho s}$ is the angle between ${\bm \rho}$ and $\bf s$. Now, we simplify the $3(N-1)$-dimensional integration into a three dimensional integration over $\theta_{\rho s}$, $\theta_{qs}$, and $s$. 
The probability to find the cluster in a hyperspherical shell in coordinate space at radius $\rho$ and in a hyperspherical shell in momentum space at radius $q$ is
\begin{eqnarray}
P_N(\rho,q)&\equiv&\rho^{3N-4}q^{3N-4}{\mathcal S_{3N-4}^2\over (2\pi)^{3(N-1)}}\int(\sin \theta_{\rho s})^{3N-5}d\theta_{\rho s} \nonumber\\ 
&\times &\int s^{3N-4}ds (\sin \theta_{qs})^{3N-5}d\theta_{qs} \nonumber\\ 
&\times &e^{-iqs\cos \theta_{qs}}R_0^*(\rho_+)R_0(\rho_-)
\label{eq.wignerProb}
\end{eqnarray}
with the normalization
\begin{eqnarray}
\int P_N(\rho,q)d\rho dq=1.
\end{eqnarray}
For the real case, we need to calculate the Wigner function numerically following Eq.~\eqref{eq.wignNN}.

The Wigner density plots for each cluster are shown in Fig.~\ref{fig.wignerall}. We see that the Wigner function can be negative due to the fact that the Wigner function is not a probability density in the classical sense, but rather a quasi-probability distribution in the phase space coordinates. 

We observe a deuteron Wigner density with a small momentum $q$ but elongated in
$\rho$ direction and to a lesser extend for $\rm ^3_\Lambda H$. For the other clusters the momenta are larger. Triton and $\rm ^3He$ have a very similar Wigner density, as expected, whereas $\rm ^4He$ is more compact. The hyper helium clusters have the largest momentum $q$ but are more compact in coordinate space as the other clusters.

\section{Summary}
\label{sec.summary}

In this paper, we presented the calculation of Wigner densities for light clusters and hyperclusters. Our approach begins with solving the Schrödinger equation using state-of-the-art potentials for nucleon–nucleon and nucleon–$\Lambda$ interactions.  For the various clusters the resulting wave functions are then projected onto hyperspherical harmonic basis states. We find that only a few  basis states are sufficient to achieve an accurate description of both the binding energy and the root-mean-square (rms) radius.

The Wigner density is obtained via a partial Fourier transform of the density matrix constructed from these hyperspherical basis states, a process that can be performed analytically.
These Wigner densities depend on both the momenta and coordinates of the nucleons, matching the degrees of freedom used in semiclassical transport models for heavy-ion collisions. This allows for a straightforward calculation of cluster formation probabilities within such transport approaches. Because this method is grounded in a realistic solution of the Schrödinger equation, it offers a path to making the coalescence model more physically accurate.

The application of this framework to extract cluster observables and multiplicities in realistic heavy-ion collision simulations will be explored in a forthcoming publication.

\begin{acknowledgments}
We acknowledge inspiring discussions ith Pengfei Zhuang. 
The work is supported by the Helmholtz Research Academy Hessen for FAIR (HFHF). 
The computational resources utilized for this work were provided by the Center for Scientific Computing (CSC) at Goethe University Frankfurt.
\end{acknowledgments}

\bibliographystyle{apsrev4-1.bst}
\bibliography{Ref}

\begin{thebibliography}{69}%
\makeatletter
\providecommand \@ifxundefined [1]{%
 \@ifx{#1\undefined}
}%
\providecommand \@ifnum [1]{%
 \ifnum #1\expandafter \@firstoftwo
 \else \expandafter \@secondoftwo
 \fi
}%
\providecommand \@ifx [1]{%
 \ifx #1\expandafter \@firstoftwo
 \else \expandafter \@secondoftwo
 \fi
}%
\providecommand \natexlab [1]{#1}%
\providecommand \enquote  [1]{``#1''}%
\providecommand \bibnamefont  [1]{#1}%
\providecommand \bibfnamefont [1]{#1}%
\providecommand \citenamefont [1]{#1}%
\providecommand \href@noop [0]{\@secondoftwo}%
\providecommand \href [0]{\begingroup \@sanitize@url \@href}%
\providecommand \@href[1]{\@@startlink{#1}\@@href}%
\providecommand \@@href[1]{\endgroup#1\@@endlink}%
\providecommand \@sanitize@url [0]{\catcode `\\12\catcode `\$12\catcode
  `\&12\catcode `\#12\catcode `\^12\catcode `\_12\catcode `\%12\relax}%
\providecommand \@@startlink[1]{}%
\providecommand \@@endlink[0]{}%
\providecommand \url  [0]{\begingroup\@sanitize@url \@url }%
\providecommand \@url [1]{\endgroup\@href {#1}{\urlprefix }}%
\providecommand \urlprefix  [0]{URL }%
\providecommand \Eprint [0]{\href }%
\providecommand \doibase [0]{http://dx.doi.org/}%
\providecommand \selectlanguage [0]{\@gobble}%
\providecommand \bibinfo  [0]{\@secondoftwo}%
\providecommand \bibfield  [0]{\@secondoftwo}%
\providecommand \translation [1]{[#1]}%
\providecommand \BibitemOpen [0]{}%
\providecommand \bibitemStop [0]{}%
\providecommand \bibitemNoStop [0]{.\EOS\space}%
\providecommand \EOS [0]{\spacefactor3000\relax}%
\providecommand \BibitemShut  [1]{\csname bibitem#1\endcsname}%
\let\auto@bib@innerbib\@empty
\bibitem [{\citenamefont
  {Schaffner-Bielich}(2008)}]{Schaffner-Bielich:2008zws}%
  \BibitemOpen
  \bibfield  {author} {\bibinfo {author} {\bibfnamefont {J.}~\bibnamefont
  {Schaffner-Bielich}},\ }\href {\doibase 10.1016/j.nuclphysa.2008.01.005}
  {\bibfield  {journal} {\bibinfo  {journal} {Nucl. Phys. A}\ }\textbf
  {\bibinfo {volume} {804}},\ \bibinfo {pages} {309} (\bibinfo {year}
  {2008})},\ \Eprint {http://arxiv.org/abs/0801.3791} {arXiv:0801.3791
  [astro-ph]} \BibitemShut {NoStop}%
\bibitem [{\citenamefont {Lattimer}\ and\ \citenamefont
  {Prakash}(2016)}]{Lattimer:2015nhk}%
  \BibitemOpen
  \bibfield  {author} {\bibinfo {author} {\bibfnamefont {J.~M.}\ \bibnamefont
  {Lattimer}}\ and\ \bibinfo {author} {\bibfnamefont {M.}~\bibnamefont
  {Prakash}},\ }\href {\doibase 10.1016/j.physrep.2015.12.005} {\bibfield
  {journal} {\bibinfo  {journal} {Phys. Rept.}\ }\textbf {\bibinfo {volume}
  {621}},\ \bibinfo {pages} {127} (\bibinfo {year} {2016})},\ \Eprint
  {http://arxiv.org/abs/1512.07820} {arXiv:1512.07820 [astro-ph.SR]}
  \BibitemShut {NoStop}%
\bibitem [{\citenamefont {Abdulhamid}\ \emph {et~al.}(2024)\citenamefont
  {Abdulhamid} \emph {et~al.}}]{STAR:2023uxk}%
  \BibitemOpen
  \bibfield  {author} {\bibinfo {author} {\bibfnamefont {M.}~\bibnamefont
  {Abdulhamid}} \emph {et~al.} (\bibinfo {collaboration} {STAR}),\ }\href
  {\doibase 10.1103/PhysRevC.110.054911} {\bibfield  {journal} {\bibinfo
  {journal} {Phys. Rev. C}\ }\textbf {\bibinfo {volume} {110}},\ \bibinfo
  {pages} {054911} (\bibinfo {year} {2024})},\ \Eprint
  {http://arxiv.org/abs/2311.11020} {arXiv:2311.11020 [nucl-ex]} \BibitemShut
  {NoStop}%
\bibitem [{\citenamefont {Abdallah}\ \emph {et~al.}(2022)\citenamefont
  {Abdallah} \emph {et~al.}}]{STAR:2021orx}%
  \BibitemOpen
  \bibfield  {author} {\bibinfo {author} {\bibfnamefont {M.}~\bibnamefont
  {Abdallah}} \emph {et~al.} (\bibinfo {collaboration} {STAR}),\ }\href
  {\doibase 10.1103/PhysRevLett.128.202301} {\bibfield  {journal} {\bibinfo
  {journal} {Phys. Rev. Lett.}\ }\textbf {\bibinfo {volume} {128}},\ \bibinfo
  {pages} {202301} (\bibinfo {year} {2022})},\ \Eprint
  {http://arxiv.org/abs/2110.09513} {arXiv:2110.09513 [nucl-ex]} \BibitemShut
  {NoStop}%
\bibitem [{\citenamefont {Acharya}\ \emph
  {et~al.}(2025{\natexlab{a}})\citenamefont {Acharya} \emph
  {et~al.}}]{ALICE:2024djx}%
  \BibitemOpen
  \bibfield  {author} {\bibinfo {author} {\bibfnamefont {S.}~\bibnamefont
  {Acharya}} \emph {et~al.} (\bibinfo {collaboration} {ALICE}),\ }\href
  {\doibase 10.1103/PhysRevLett.134.162301} {\bibfield  {journal} {\bibinfo
  {journal} {Phys. Rev. Lett.}\ }\textbf {\bibinfo {volume} {134}},\ \bibinfo
  {pages} {162301} (\bibinfo {year} {2025}{\natexlab{a}})},\ \Eprint
  {http://arxiv.org/abs/2410.17769} {arXiv:2410.17769 [nucl-ex]} \BibitemShut
  {NoStop}%
\bibitem [{\citenamefont {Acharya}\ \emph {et~al.}(2023)\citenamefont {Acharya}
  \emph {et~al.}}]{ALICE:2022sco}%
  \BibitemOpen
  \bibfield  {author} {\bibinfo {author} {\bibfnamefont {S.}~\bibnamefont
  {Acharya}} \emph {et~al.} (\bibinfo {collaboration} {ALICE}),\ }\href
  {\doibase 10.1103/PhysRevLett.131.102302} {\bibfield  {journal} {\bibinfo
  {journal} {Phys. Rev. Lett.}\ }\textbf {\bibinfo {volume} {131}},\ \bibinfo
  {pages} {102302} (\bibinfo {year} {2023})},\ \Eprint
  {http://arxiv.org/abs/2209.07360} {arXiv:2209.07360 [nucl-ex]} \BibitemShut
  {NoStop}%
\bibitem [{\citenamefont {Acharya}\ \emph
  {et~al.}(2025{\natexlab{b}})\citenamefont {Acharya} \emph
  {et~al.}}]{ALICE:2024koa}%
  \BibitemOpen
  \bibfield  {author} {\bibinfo {author} {\bibfnamefont {S.}~\bibnamefont
  {Acharya}} \emph {et~al.} (\bibinfo {collaboration} {ALICE}),\ }\href
  {\doibase 10.1016/j.physletb.2024.139066} {\bibfield  {journal} {\bibinfo
  {journal} {Phys. Lett. B}\ }\textbf {\bibinfo {volume} {860}},\ \bibinfo
  {pages} {139066} (\bibinfo {year} {2025}{\natexlab{b}})},\ \Eprint
  {http://arxiv.org/abs/2405.19839} {arXiv:2405.19839 [nucl-ex]} \BibitemShut
  {NoStop}%
\bibitem [{\citenamefont {Oliinychenko}\ \emph {et~al.}(2019)\citenamefont
  {Oliinychenko}, \citenamefont {Pang}, \citenamefont {Elfner},\ and\
  \citenamefont {Koch}}]{Oliinychenko:2018ugs}%
  \BibitemOpen
  \bibfield  {author} {\bibinfo {author} {\bibfnamefont {D.}~\bibnamefont
  {Oliinychenko}}, \bibinfo {author} {\bibfnamefont {L.-G.}\ \bibnamefont
  {Pang}}, \bibinfo {author} {\bibfnamefont {H.}~\bibnamefont {Elfner}}, \ and\
  \bibinfo {author} {\bibfnamefont {V.}~\bibnamefont {Koch}} (\bibinfo
  {collaboration} {SMASH}),\ }\href {\doibase 10.1103/PhysRevC.99.044907}
  {\bibfield  {journal} {\bibinfo  {journal} {Phys. Rev. C}\ }\textbf {\bibinfo
  {volume} {99}},\ \bibinfo {pages} {044907} (\bibinfo {year} {2019})},\
  \Eprint {http://arxiv.org/abs/1809.03071} {arXiv:1809.03071 [hep-ph]}
  \BibitemShut {NoStop}%
\bibitem [{\citenamefont {Oliinychenko}\ \emph {et~al.}(2021)\citenamefont
  {Oliinychenko}, \citenamefont {Shen},\ and\ \citenamefont
  {Koch}}]{Oliinychenko:2020znl}%
  \BibitemOpen
  \bibfield  {author} {\bibinfo {author} {\bibfnamefont {D.}~\bibnamefont
  {Oliinychenko}}, \bibinfo {author} {\bibfnamefont {C.}~\bibnamefont {Shen}},
  \ and\ \bibinfo {author} {\bibfnamefont {V.}~\bibnamefont {Koch}} (\bibinfo
  {collaboration} {SMASH}),\ }\href {\doibase 10.1103/PhysRevC.103.034913}
  {\bibfield  {journal} {\bibinfo  {journal} {Phys. Rev. C}\ }\textbf {\bibinfo
  {volume} {103}},\ \bibinfo {pages} {034913} (\bibinfo {year} {2021})},\
  \Eprint {http://arxiv.org/abs/2009.01915} {arXiv:2009.01915 [hep-ph]}
  \BibitemShut {NoStop}%
\bibitem [{\citenamefont {Aichelin}(1991)}]{Aichelin:1991xy}%
  \BibitemOpen
  \bibfield  {author} {\bibinfo {author} {\bibfnamefont {J.}~\bibnamefont
  {Aichelin}},\ }\href {\doibase 10.1016/0370-1573(91)90094-3} {\bibfield
  {journal} {\bibinfo  {journal} {Phys. Rept.}\ }\textbf {\bibinfo {volume}
  {202}},\ \bibinfo {pages} {233} (\bibinfo {year} {1991})}\BibitemShut
  {NoStop}%
\bibitem [{\citenamefont {Aichelin}\ \emph {et~al.}(2020)\citenamefont
  {Aichelin}, \citenamefont {Bratkovskaya}, \citenamefont {Le~F{\`e}vre},
  \citenamefont {Kireyeu}, \citenamefont {Kolesnikov}, \citenamefont {Leifels},
  \citenamefont {Voronyuk},\ and\ \citenamefont {Coci}}]{Aichelin:2019tnk}%
  \BibitemOpen
  \bibfield  {author} {\bibinfo {author} {\bibfnamefont {J.}~\bibnamefont
  {Aichelin}}, \bibinfo {author} {\bibfnamefont {E.}~\bibnamefont
  {Bratkovskaya}}, \bibinfo {author} {\bibfnamefont {A.}~\bibnamefont
  {Le~F{\`e}vre}}, \bibinfo {author} {\bibfnamefont {V.}~\bibnamefont
  {Kireyeu}}, \bibinfo {author} {\bibfnamefont {V.}~\bibnamefont {Kolesnikov}},
  \bibinfo {author} {\bibfnamefont {Y.}~\bibnamefont {Leifels}}, \bibinfo
  {author} {\bibfnamefont {V.}~\bibnamefont {Voronyuk}}, \ and\ \bibinfo
  {author} {\bibfnamefont {G.}~\bibnamefont {Coci}},\ }\href {\doibase
  10.1103/PhysRevC.101.044905} {\bibfield  {journal} {\bibinfo  {journal}
  {Phys. Rev. C}\ }\textbf {\bibinfo {volume} {101}},\ \bibinfo {pages}
  {044905} (\bibinfo {year} {2020})},\ \Eprint
  {http://arxiv.org/abs/1907.03860} {arXiv:1907.03860 [nucl-th]} \BibitemShut
  {NoStop}%
\bibitem [{\citenamefont {Gl{\"a}{\ss}el}\ \emph {et~al.}(2022)\citenamefont
  {Gl{\"a}{\ss}el}, \citenamefont {Kireyeu}, \citenamefont {Voronyuk},
  \citenamefont {Aichelin}, \citenamefont {Blume}, \citenamefont
  {Bratkovskaya}, \citenamefont {Coci}, \citenamefont {Kolesnikov},\ and\
  \citenamefont {Winn}}]{Glassel:2021rod}%
  \BibitemOpen
  \bibfield  {author} {\bibinfo {author} {\bibfnamefont {S.}~\bibnamefont
  {Gl{\"a}{\ss}el}}, \bibinfo {author} {\bibfnamefont {V.}~\bibnamefont
  {Kireyeu}}, \bibinfo {author} {\bibfnamefont {V.}~\bibnamefont {Voronyuk}},
  \bibinfo {author} {\bibfnamefont {J.}~\bibnamefont {Aichelin}}, \bibinfo
  {author} {\bibfnamefont {C.}~\bibnamefont {Blume}}, \bibinfo {author}
  {\bibfnamefont {E.}~\bibnamefont {Bratkovskaya}}, \bibinfo {author}
  {\bibfnamefont {G.}~\bibnamefont {Coci}}, \bibinfo {author} {\bibfnamefont
  {V.}~\bibnamefont {Kolesnikov}}, \ and\ \bibinfo {author} {\bibfnamefont
  {M.}~\bibnamefont {Winn}},\ }\href {\doibase 10.1103/PhysRevC.105.014908}
  {\bibfield  {journal} {\bibinfo  {journal} {Phys. Rev. C}\ }\textbf {\bibinfo
  {volume} {105}},\ \bibinfo {pages} {014908} (\bibinfo {year} {2022})},\
  \Eprint {http://arxiv.org/abs/2106.14839} {arXiv:2106.14839 [nucl-th]}
  \BibitemShut {NoStop}%
\bibitem [{\citenamefont {Coci}\ \emph {et~al.}(2023)\citenamefont {Coci},
  \citenamefont {Gl{\"a}{\ss}el}, \citenamefont {Kireyeu}, \citenamefont
  {Aichelin}, \citenamefont {Blume}, \citenamefont {Bratkovskaya},
  \citenamefont {Kolesnikov},\ and\ \citenamefont {Voronyuk}}]{Coci:2023daq}%
  \BibitemOpen
  \bibfield  {author} {\bibinfo {author} {\bibfnamefont {G.}~\bibnamefont
  {Coci}}, \bibinfo {author} {\bibfnamefont {S.}~\bibnamefont
  {Gl{\"a}{\ss}el}}, \bibinfo {author} {\bibfnamefont {V.}~\bibnamefont
  {Kireyeu}}, \bibinfo {author} {\bibfnamefont {J.}~\bibnamefont {Aichelin}},
  \bibinfo {author} {\bibfnamefont {C.}~\bibnamefont {Blume}}, \bibinfo
  {author} {\bibfnamefont {E.}~\bibnamefont {Bratkovskaya}}, \bibinfo {author}
  {\bibfnamefont {V.}~\bibnamefont {Kolesnikov}}, \ and\ \bibinfo {author}
  {\bibfnamefont {V.}~\bibnamefont {Voronyuk}},\ }\href {\doibase
  10.1103/PhysRevC.108.014902} {\bibfield  {journal} {\bibinfo  {journal}
  {Phys. Rev. C}\ }\textbf {\bibinfo {volume} {108}},\ \bibinfo {pages}
  {014902} (\bibinfo {year} {2023})},\ \Eprint
  {http://arxiv.org/abs/2303.02279} {arXiv:2303.02279 [nucl-th]} \BibitemShut
  {NoStop}%
\bibitem [{\citenamefont {Kireyeu}\ \emph {et~al.}(2024)\citenamefont
  {Kireyeu}, \citenamefont {Voronyuk}, \citenamefont {Winn}, \citenamefont
  {Gl{\"a}{\ss}el}, \citenamefont {Aichelin}, \citenamefont {Blume},
  \citenamefont {Bratkovskaya}, \citenamefont {Coci},\ and\ \citenamefont
  {Zhao}}]{Kireyeu:2024hjo}%
  \BibitemOpen
  \bibfield  {author} {\bibinfo {author} {\bibfnamefont {V.}~\bibnamefont
  {Kireyeu}}, \bibinfo {author} {\bibfnamefont {V.}~\bibnamefont {Voronyuk}},
  \bibinfo {author} {\bibfnamefont {M.}~\bibnamefont {Winn}}, \bibinfo {author}
  {\bibfnamefont {S.}~\bibnamefont {Gl{\"a}{\ss}el}}, \bibinfo {author}
  {\bibfnamefont {J.}~\bibnamefont {Aichelin}}, \bibinfo {author}
  {\bibfnamefont {C.}~\bibnamefont {Blume}}, \bibinfo {author} {\bibfnamefont
  {E.}~\bibnamefont {Bratkovskaya}}, \bibinfo {author} {\bibfnamefont
  {G.}~\bibnamefont {Coci}}, \ and\ \bibinfo {author} {\bibfnamefont
  {J.}~\bibnamefont {Zhao}},\ }\href@noop {} {\  (\bibinfo {year} {2024})},\
  \Eprint {http://arxiv.org/abs/2411.04969} {arXiv:2411.04969 [nucl-th]}
  \BibitemShut {NoStop}%
\bibitem [{\citenamefont {Andronic}\ \emph {et~al.}(2011)\citenamefont
  {Andronic}, \citenamefont {Braun-Munzinger}, \citenamefont {Stachel},\ and\
  \citenamefont {Stocker}}]{Andronic:2010qu}%
  \BibitemOpen
  \bibfield  {author} {\bibinfo {author} {\bibfnamefont {A.}~\bibnamefont
  {Andronic}}, \bibinfo {author} {\bibfnamefont {P.}~\bibnamefont
  {Braun-Munzinger}}, \bibinfo {author} {\bibfnamefont {J.}~\bibnamefont
  {Stachel}}, \ and\ \bibinfo {author} {\bibfnamefont {H.}~\bibnamefont
  {Stocker}},\ }\href {\doibase 10.1016/j.physletb.2011.01.053} {\bibfield
  {journal} {\bibinfo  {journal} {Phys. Lett. B}\ }\textbf {\bibinfo {volume}
  {697}},\ \bibinfo {pages} {203} (\bibinfo {year} {2011})},\ \Eprint
  {http://arxiv.org/abs/1010.2995} {arXiv:1010.2995 [nucl-th]} \BibitemShut
  {NoStop}%
\bibitem [{\citenamefont {Steinheimer}\ \emph {et~al.}(2012)\citenamefont
  {Steinheimer}, \citenamefont {Gudima}, \citenamefont {Botvina}, \citenamefont
  {Mishustin}, \citenamefont {Bleicher},\ and\ \citenamefont
  {Stocker}}]{Steinheimer:2012tb}%
  \BibitemOpen
  \bibfield  {author} {\bibinfo {author} {\bibfnamefont {J.}~\bibnamefont
  {Steinheimer}}, \bibinfo {author} {\bibfnamefont {K.}~\bibnamefont {Gudima}},
  \bibinfo {author} {\bibfnamefont {A.}~\bibnamefont {Botvina}}, \bibinfo
  {author} {\bibfnamefont {I.}~\bibnamefont {Mishustin}}, \bibinfo {author}
  {\bibfnamefont {M.}~\bibnamefont {Bleicher}}, \ and\ \bibinfo {author}
  {\bibfnamefont {H.}~\bibnamefont {Stocker}},\ }\href {\doibase
  10.1016/j.physletb.2012.06.069} {\bibfield  {journal} {\bibinfo  {journal}
  {Phys. Lett. B}\ }\textbf {\bibinfo {volume} {714}},\ \bibinfo {pages} {85}
  (\bibinfo {year} {2012})},\ \Eprint {http://arxiv.org/abs/1203.2547}
  {arXiv:1203.2547 [nucl-th]} \BibitemShut {NoStop}%
\bibitem [{\citenamefont {Sombun}\ \emph {et~al.}(2019)\citenamefont {Sombun},
  \citenamefont {Tomuang}, \citenamefont {Limphirat}, \citenamefont {Hillmann},
  \citenamefont {Herold}, \citenamefont {Steinheimer}, \citenamefont {Yan},\
  and\ \citenamefont {Bleicher}}]{Sombun:2018yqh}%
  \BibitemOpen
  \bibfield  {author} {\bibinfo {author} {\bibfnamefont {S.}~\bibnamefont
  {Sombun}}, \bibinfo {author} {\bibfnamefont {K.}~\bibnamefont {Tomuang}},
  \bibinfo {author} {\bibfnamefont {A.}~\bibnamefont {Limphirat}}, \bibinfo
  {author} {\bibfnamefont {P.}~\bibnamefont {Hillmann}}, \bibinfo {author}
  {\bibfnamefont {C.}~\bibnamefont {Herold}}, \bibinfo {author} {\bibfnamefont
  {J.}~\bibnamefont {Steinheimer}}, \bibinfo {author} {\bibfnamefont
  {Y.}~\bibnamefont {Yan}}, \ and\ \bibinfo {author} {\bibfnamefont
  {M.}~\bibnamefont {Bleicher}},\ }\href {\doibase 10.1103/PhysRevC.99.014901}
  {\bibfield  {journal} {\bibinfo  {journal} {Phys. Rev. C}\ }\textbf {\bibinfo
  {volume} {99}},\ \bibinfo {pages} {014901} (\bibinfo {year} {2019})},\
  \Eprint {http://arxiv.org/abs/1805.11509} {arXiv:1805.11509 [nucl-th]}
  \BibitemShut {NoStop}%
\bibitem [{\citenamefont {Liu}\ \emph {et~al.}(2020)\citenamefont {Liu},
  \citenamefont {Zhang}, \citenamefont {He}, \citenamefont {Sun}, \citenamefont
  {Yu},\ and\ \citenamefont {Luo}}]{Liu:2019nii}%
  \BibitemOpen
  \bibfield  {author} {\bibinfo {author} {\bibfnamefont {H.}~\bibnamefont
  {Liu}}, \bibinfo {author} {\bibfnamefont {D.}~\bibnamefont {Zhang}}, \bibinfo
  {author} {\bibfnamefont {S.}~\bibnamefont {He}}, \bibinfo {author}
  {\bibfnamefont {K.-j.}\ \bibnamefont {Sun}}, \bibinfo {author} {\bibfnamefont
  {N.}~\bibnamefont {Yu}}, \ and\ \bibinfo {author} {\bibfnamefont
  {X.}~\bibnamefont {Luo}},\ }\href {\doibase 10.1016/j.physletb.2020.135452}
  {\bibfield  {journal} {\bibinfo  {journal} {Phys. Lett. B}\ }\textbf
  {\bibinfo {volume} {805}},\ \bibinfo {pages} {135452} (\bibinfo {year}
  {2020})},\ \bibinfo {note} {[Erratum: Phys.Lett.B 829, 137132 (2022)]},\
  \Eprint {http://arxiv.org/abs/1909.09304} {arXiv:1909.09304 [nucl-th]}
  \BibitemShut {NoStop}%
\bibitem [{\citenamefont {Kireyeu}\ \emph {et~al.}(2022)\citenamefont
  {Kireyeu}, \citenamefont {Steinheimer}, \citenamefont {Aichelin},
  \citenamefont {Bleicher},\ and\ \citenamefont
  {Bratkovskaya}}]{Kireyeu:2022qmv}%
  \BibitemOpen
  \bibfield  {author} {\bibinfo {author} {\bibfnamefont {V.}~\bibnamefont
  {Kireyeu}}, \bibinfo {author} {\bibfnamefont {J.}~\bibnamefont
  {Steinheimer}}, \bibinfo {author} {\bibfnamefont {J.}~\bibnamefont
  {Aichelin}}, \bibinfo {author} {\bibfnamefont {M.}~\bibnamefont {Bleicher}},
  \ and\ \bibinfo {author} {\bibfnamefont {E.}~\bibnamefont {Bratkovskaya}},\
  }\href {\doibase 10.1103/PhysRevC.105.044909} {\bibfield  {journal} {\bibinfo
   {journal} {Phys. Rev. C}\ }\textbf {\bibinfo {volume} {105}},\ \bibinfo
  {pages} {044909} (\bibinfo {year} {2022})},\ \Eprint
  {http://arxiv.org/abs/2201.13374} {arXiv:2201.13374 [nucl-th]} \BibitemShut
  {NoStop}%
\bibitem [{\citenamefont {Reichert}\ \emph {et~al.}(2025)\citenamefont
  {Reichert}, \citenamefont {Omana~Kuttan}, \citenamefont {Kittiratpattana},
  \citenamefont {Buyukcizmeci}, \citenamefont {Botvina}, \citenamefont
  {Steinheimer},\ and\ \citenamefont {Bleicher}}]{Reichert:2025rnw}%
  \BibitemOpen
  \bibfield  {author} {\bibinfo {author} {\bibfnamefont {T.}~\bibnamefont
  {Reichert}}, \bibinfo {author} {\bibfnamefont {M.}~\bibnamefont
  {Omana~Kuttan}}, \bibinfo {author} {\bibfnamefont {A.}~\bibnamefont
  {Kittiratpattana}}, \bibinfo {author} {\bibfnamefont {N.}~\bibnamefont
  {Buyukcizmeci}}, \bibinfo {author} {\bibfnamefont {A.}~\bibnamefont
  {Botvina}}, \bibinfo {author} {\bibfnamefont {J.}~\bibnamefont
  {Steinheimer}}, \ and\ \bibinfo {author} {\bibfnamefont {M.}~\bibnamefont
  {Bleicher}},\ }\href@noop {} {\  (\bibinfo {year} {2025})},\ \Eprint
  {http://arxiv.org/abs/2504.17389} {arXiv:2504.17389 [nucl-th]} \BibitemShut
  {NoStop}%
\bibitem [{\citenamefont {Scheibl}\ and\ \citenamefont
  {Heinz}(1999)}]{Scheibl:1998tk}%
  \BibitemOpen
  \bibfield  {author} {\bibinfo {author} {\bibfnamefont {R.}~\bibnamefont
  {Scheibl}}\ and\ \bibinfo {author} {\bibfnamefont {U.~W.}\ \bibnamefont
  {Heinz}},\ }\href {\doibase 10.1103/PhysRevC.59.1585} {\bibfield  {journal}
  {\bibinfo  {journal} {Phys. Rev. C}\ }\textbf {\bibinfo {volume} {59}},\
  \bibinfo {pages} {1585} (\bibinfo {year} {1999})},\ \Eprint
  {http://arxiv.org/abs/nucl-th/9809092} {arXiv:nucl-th/9809092} \BibitemShut
  {NoStop}%
\bibitem [{\citenamefont {Zhu}\ \emph {et~al.}(2015)\citenamefont {Zhu},
  \citenamefont {Ko},\ and\ \citenamefont {Yin}}]{Zhu:2015voa}%
  \BibitemOpen
  \bibfield  {author} {\bibinfo {author} {\bibfnamefont {L.}~\bibnamefont
  {Zhu}}, \bibinfo {author} {\bibfnamefont {C.~M.}\ \bibnamefont {Ko}}, \ and\
  \bibinfo {author} {\bibfnamefont {X.}~\bibnamefont {Yin}},\ }\href {\doibase
  10.1103/PhysRevC.92.064911} {\bibfield  {journal} {\bibinfo  {journal} {Phys.
  Rev. C}\ }\textbf {\bibinfo {volume} {92}},\ \bibinfo {pages} {064911}
  (\bibinfo {year} {2015})},\ \Eprint {http://arxiv.org/abs/1510.03568}
  {arXiv:1510.03568 [nucl-th]} \BibitemShut {NoStop}%
\bibitem [{\citenamefont {Aichelin}\ and\ \citenamefont
  {Remler}(1987)}]{Aichelin:1987rh}%
  \BibitemOpen
  \bibfield  {author} {\bibinfo {author} {\bibfnamefont {J.}~\bibnamefont
  {Aichelin}}\ and\ \bibinfo {author} {\bibfnamefont {E.~A.}\ \bibnamefont
  {Remler}},\ }\href {\doibase 10.1103/PhysRevC.35.1291} {\bibfield  {journal}
  {\bibinfo  {journal} {Phys. Rev. C}\ }\textbf {\bibinfo {volume} {35}},\
  \bibinfo {pages} {1291} (\bibinfo {year} {1987})}\BibitemShut {NoStop}%
\bibitem [{\citenamefont {Sun}\ and\ \citenamefont {Ko}(2021)}]{Sun:2020uoj}%
  \BibitemOpen
  \bibfield  {author} {\bibinfo {author} {\bibfnamefont {K.-J.}\ \bibnamefont
  {Sun}}\ and\ \bibinfo {author} {\bibfnamefont {C.~M.}\ \bibnamefont {Ko}},\
  }\href {\doibase 10.1103/PhysRevC.103.064909} {\bibfield  {journal} {\bibinfo
   {journal} {Phys. Rev. C}\ }\textbf {\bibinfo {volume} {103}},\ \bibinfo
  {pages} {064909} (\bibinfo {year} {2021})},\ \Eprint
  {http://arxiv.org/abs/2005.00182} {arXiv:2005.00182 [nucl-th]} \BibitemShut
  {NoStop}%
\bibitem [{\citenamefont {Wang}\ \emph {et~al.}(2025)\citenamefont {Wang},
  \citenamefont {Song}, \citenamefont {Wu}, \citenamefont {Xue},\ and\
  \citenamefont {Shao}}]{Wang:2025ify}%
  \BibitemOpen
  \bibfield  {author} {\bibinfo {author} {\bibfnamefont {R.-Q.}\ \bibnamefont
  {Wang}}, \bibinfo {author} {\bibfnamefont {J.}~\bibnamefont {Song}}, \bibinfo
  {author} {\bibfnamefont {M.-Y.}\ \bibnamefont {Wu}}, \bibinfo {author}
  {\bibfnamefont {H.-T.}\ \bibnamefont {Xue}}, \ and\ \bibinfo {author}
  {\bibfnamefont {F.-L.}\ \bibnamefont {Shao}},\ }\href@noop {} {\  (\bibinfo
  {year} {2025})},\ \Eprint {http://arxiv.org/abs/2504.13640} {arXiv:2504.13640
  [nucl-th]} \BibitemShut {NoStop}%
\bibitem [{\citenamefont {Wang}\ \emph {et~al.}(2024)\citenamefont {Wang},
  \citenamefont {Li}, \citenamefont {Song},\ and\ \citenamefont
  {Shao}}]{Wang:2023rpd}%
  \BibitemOpen
  \bibfield  {author} {\bibinfo {author} {\bibfnamefont {R.-Q.}\ \bibnamefont
  {Wang}}, \bibinfo {author} {\bibfnamefont {Y.-H.}\ \bibnamefont {Li}},
  \bibinfo {author} {\bibfnamefont {J.}~\bibnamefont {Song}}, \ and\ \bibinfo
  {author} {\bibfnamefont {F.-L.}\ \bibnamefont {Shao}},\ }\href {\doibase
  10.1103/PhysRevC.109.034907} {\bibfield  {journal} {\bibinfo  {journal}
  {Phys. Rev. C}\ }\textbf {\bibinfo {volume} {109}},\ \bibinfo {pages}
  {034907} (\bibinfo {year} {2024})},\ \Eprint
  {http://arxiv.org/abs/2309.16296} {arXiv:2309.16296 [nucl-th]} \BibitemShut
  {NoStop}%
\bibitem [{\citenamefont {Liu}\ \emph {et~al.}(2024{\natexlab{a}})\citenamefont
  {Liu}, \citenamefont {Ko}, \citenamefont {Ma}, \citenamefont {Mazzaschi},
  \citenamefont {Puccio}, \citenamefont {Shou}, \citenamefont {Sun},\ and\
  \citenamefont {Wang}}]{Liu:2024ygk}%
  \BibitemOpen
  \bibfield  {author} {\bibinfo {author} {\bibfnamefont {D.-N.}\ \bibnamefont
  {Liu}}, \bibinfo {author} {\bibfnamefont {C.~M.}\ \bibnamefont {Ko}},
  \bibinfo {author} {\bibfnamefont {Y.-G.}\ \bibnamefont {Ma}}, \bibinfo
  {author} {\bibfnamefont {F.}~\bibnamefont {Mazzaschi}}, \bibinfo {author}
  {\bibfnamefont {M.}~\bibnamefont {Puccio}}, \bibinfo {author} {\bibfnamefont
  {Q.-Y.}\ \bibnamefont {Shou}}, \bibinfo {author} {\bibfnamefont {K.-J.}\
  \bibnamefont {Sun}}, \ and\ \bibinfo {author} {\bibfnamefont {Y.-Z.}\
  \bibnamefont {Wang}},\ }\href {\doibase 10.1016/j.physletb.2024.138855}
  {\bibfield  {journal} {\bibinfo  {journal} {Phys. Lett. B}\ }\textbf
  {\bibinfo {volume} {855}},\ \bibinfo {pages} {138855} (\bibinfo {year}
  {2024}{\natexlab{a}})},\ \Eprint {http://arxiv.org/abs/2404.02701}
  {arXiv:2404.02701 [nucl-th]} \BibitemShut {NoStop}%
\bibitem [{\citenamefont {Lin}\ \emph {et~al.}(2025)\citenamefont {Lin},
  \citenamefont {Huang}, \citenamefont {Pang}, \citenamefont {Luo},\ and\
  \citenamefont {Wang}}]{Lin:2025dsm}%
  \BibitemOpen
  \bibfield  {author} {\bibinfo {author} {\bibfnamefont {Q.-R.}\ \bibnamefont
  {Lin}}, \bibinfo {author} {\bibfnamefont {Y.-J.}\ \bibnamefont {Huang}},
  \bibinfo {author} {\bibfnamefont {L.-G.}\ \bibnamefont {Pang}}, \bibinfo
  {author} {\bibfnamefont {X.-F.}\ \bibnamefont {Luo}}, \ and\ \bibinfo
  {author} {\bibfnamefont {X.-N.}\ \bibnamefont {Wang}},\ }\href@noop {} {\
  (\bibinfo {year} {2025})},\ \Eprint {http://arxiv.org/abs/2503.01128}
  {arXiv:2503.01128 [hep-ph]} \BibitemShut {NoStop}%
\bibitem [{\citenamefont {Bellini}\ \emph {et~al.}(2021)\citenamefont
  {Bellini}, \citenamefont {Blum}, \citenamefont {Kalweit},\ and\ \citenamefont
  {Puccio}}]{Bellini:2020cbj}%
  \BibitemOpen
  \bibfield  {author} {\bibinfo {author} {\bibfnamefont {F.}~\bibnamefont
  {Bellini}}, \bibinfo {author} {\bibfnamefont {K.}~\bibnamefont {Blum}},
  \bibinfo {author} {\bibfnamefont {A.~P.}\ \bibnamefont {Kalweit}}, \ and\
  \bibinfo {author} {\bibfnamefont {M.}~\bibnamefont {Puccio}},\ }\href
  {\doibase 10.1103/PhysRevC.103.014907} {\bibfield  {journal} {\bibinfo
  {journal} {Phys. Rev. C}\ }\textbf {\bibinfo {volume} {103}},\ \bibinfo
  {pages} {014907} (\bibinfo {year} {2021})},\ \Eprint
  {http://arxiv.org/abs/2007.01750} {arXiv:2007.01750 [nucl-th]} \BibitemShut
  {NoStop}%
\bibitem [{\citenamefont {Mahlein}\ \emph {et~al.}(2023)\citenamefont
  {Mahlein}, \citenamefont {Barioglio}, \citenamefont {Bellini}, \citenamefont
  {Fabbietti}, \citenamefont {Pinto}, \citenamefont {Singh},\ and\
  \citenamefont {Tripathy}}]{Mahlein:2023fmx}%
  \BibitemOpen
  \bibfield  {author} {\bibinfo {author} {\bibfnamefont {M.}~\bibnamefont
  {Mahlein}}, \bibinfo {author} {\bibfnamefont {L.}~\bibnamefont {Barioglio}},
  \bibinfo {author} {\bibfnamefont {F.}~\bibnamefont {Bellini}}, \bibinfo
  {author} {\bibfnamefont {L.}~\bibnamefont {Fabbietti}}, \bibinfo {author}
  {\bibfnamefont {C.}~\bibnamefont {Pinto}}, \bibinfo {author} {\bibfnamefont
  {B.}~\bibnamefont {Singh}}, \ and\ \bibinfo {author} {\bibfnamefont
  {S.}~\bibnamefont {Tripathy}},\ }\href {\doibase
  10.1140/epjc/s10052-023-11972-3} {\bibfield  {journal} {\bibinfo  {journal}
  {Eur. Phys. J. C}\ }\textbf {\bibinfo {volume} {83}},\ \bibinfo {pages} {804}
  (\bibinfo {year} {2023})},\ \Eprint {http://arxiv.org/abs/2302.12696}
  {arXiv:2302.12696 [hep-ex]} \BibitemShut {NoStop}%
\bibitem [{\citenamefont {Barnea}\ \emph {et~al.}(2000)\citenamefont {Barnea},
  \citenamefont {Leidemann},\ and\ \citenamefont {Orlandini}}]{Barnea:1999be}%
  \BibitemOpen
  \bibfield  {author} {\bibinfo {author} {\bibfnamefont {N.}~\bibnamefont
  {Barnea}}, \bibinfo {author} {\bibfnamefont {W.}~\bibnamefont {Leidemann}}, \
  and\ \bibinfo {author} {\bibfnamefont {G.}~\bibnamefont {Orlandini}},\ }\href
  {\doibase 10.1103/PhysRevC.61.054001} {\bibfield  {journal} {\bibinfo
  {journal} {Phys. Rev. C}\ }\textbf {\bibinfo {volume} {61}},\ \bibinfo
  {pages} {054001} (\bibinfo {year} {2000})},\ \Eprint
  {http://arxiv.org/abs/nucl-th/9910062} {arXiv:nucl-th/9910062} \BibitemShut
  {NoStop}%
\bibitem [{\citenamefont {Marcucci}\ \emph {et~al.}(2020)\citenamefont
  {Marcucci}, \citenamefont {Dohet-Eraly}, \citenamefont {Girlanda},
  \citenamefont {Gnech}, \citenamefont {Kievsky},\ and\ \citenamefont
  {Viviani}}]{Marcucci:2019hml}%
  \BibitemOpen
  \bibfield  {author} {\bibinfo {author} {\bibfnamefont {L.~E.}\ \bibnamefont
  {Marcucci}}, \bibinfo {author} {\bibfnamefont {J.}~\bibnamefont
  {Dohet-Eraly}}, \bibinfo {author} {\bibfnamefont {L.}~\bibnamefont
  {Girlanda}}, \bibinfo {author} {\bibfnamefont {A.}~\bibnamefont {Gnech}},
  \bibinfo {author} {\bibfnamefont {A.}~\bibnamefont {Kievsky}}, \ and\
  \bibinfo {author} {\bibfnamefont {M.}~\bibnamefont {Viviani}},\ }\href
  {\doibase 10.3389/fphy.2020.00069} {\bibfield  {journal} {\bibinfo  {journal}
  {Front. in Phys.}\ }\textbf {\bibinfo {volume} {8}},\ \bibinfo {pages} {69}
  (\bibinfo {year} {2020})},\ \Eprint {http://arxiv.org/abs/1912.09751}
  {arXiv:1912.09751 [nucl-th]} \BibitemShut {NoStop}%
\bibitem [{\citenamefont {Zhao}\ \emph {et~al.}(2020)\citenamefont {Zhao},
  \citenamefont {Shi},\ and\ \citenamefont {Zhuang}}]{Zhao:2020nwy}%
  \BibitemOpen
  \bibfield  {author} {\bibinfo {author} {\bibfnamefont {J.}~\bibnamefont
  {Zhao}}, \bibinfo {author} {\bibfnamefont {S.}~\bibnamefont {Shi}}, \ and\
  \bibinfo {author} {\bibfnamefont {P.}~\bibnamefont {Zhuang}},\ }\href
  {\doibase 10.1103/PhysRevD.102.114001} {\bibfield  {journal} {\bibinfo
  {journal} {Phys. Rev. D}\ }\textbf {\bibinfo {volume} {102}},\ \bibinfo
  {pages} {114001} (\bibinfo {year} {2020})},\ \Eprint
  {http://arxiv.org/abs/2009.10319} {arXiv:2009.10319 [hep-ph]} \BibitemShut
  {NoStop}%
\bibitem [{\citenamefont {Kohn}(1948)}]{Kohn:1948col}%
  \BibitemOpen
  \bibfield  {author} {\bibinfo {author} {\bibfnamefont {W.}~\bibnamefont
  {Kohn}},\ }\href {\doibase 10.1103/PhysRev.74.1763} {\bibfield  {journal}
  {\bibinfo  {journal} {Phys. Rev.}\ }\textbf {\bibinfo {volume} {74}},\
  \bibinfo {pages} {1763} (\bibinfo {year} {1948})}\BibitemShut {NoStop}%
\bibitem [{\citenamefont {Demkov}\ and\ \citenamefont
  {Kemmer}(1963)}]{demkov1963variational}%
  \BibitemOpen
  \bibfield  {author} {\bibinfo {author} {\bibfnamefont {I.}~\bibnamefont
  {Demkov}}\ and\ \bibinfo {author} {\bibfnamefont {N.}~\bibnamefont
  {Kemmer}},\ }\href@noop {} {\bibfield  {journal} {\bibinfo  {journal} {(No
  Title)}\ } (\bibinfo {year} {1963})}\BibitemShut {NoStop}%
\bibitem [{\citenamefont {Kamimura}(1988)}]{Kamimura:1988zz}%
  \BibitemOpen
  \bibfield  {author} {\bibinfo {author} {\bibfnamefont {M.}~\bibnamefont
  {Kamimura}},\ }\href {\doibase 10.1103/PhysRevA.38.621} {\bibfield  {journal}
  {\bibinfo  {journal} {Phys. Rev. A}\ }\textbf {\bibinfo {volume} {38}},\
  \bibinfo {pages} {621} (\bibinfo {year} {1988})}\BibitemShut {NoStop}%
\bibitem [{\citenamefont {Faddeev}(1961)}]{Faddeev:1960su}%
  \BibitemOpen
  \bibfield  {author} {\bibinfo {author} {\bibfnamefont {L.~D.}\ \bibnamefont
  {Faddeev}},\ }\href@noop {} {\bibfield  {journal} {\bibinfo  {journal} {Sov.
  Phys. JETP}\ }\textbf {\bibinfo {volume} {12}},\ \bibinfo {pages} {1014}
  (\bibinfo {year} {1961})}\BibitemShut {NoStop}%
\bibitem [{\citenamefont {Gl{\"o}ckle}(2012)}]{glockle2012quantum}%
  \BibitemOpen
  \bibfield  {author} {\bibinfo {author} {\bibfnamefont {W.}~\bibnamefont
  {Gl{\"o}ckle}},\ }\href@noop {} {\emph {\bibinfo {title} {The quantum
  mechanical few-body problem}}}\ (\bibinfo  {publisher} {Springer Science \&
  Business Media},\ \bibinfo {year} {2012})\BibitemShut {NoStop}%
\bibitem [{\citenamefont {Leidemann}\ and\ \citenamefont
  {Orlandini}(2013)}]{Leidemann:2012hr}%
  \BibitemOpen
  \bibfield  {author} {\bibinfo {author} {\bibfnamefont {W.}~\bibnamefont
  {Leidemann}}\ and\ \bibinfo {author} {\bibfnamefont {G.}~\bibnamefont
  {Orlandini}},\ }\href {\doibase 10.1016/j.ppnp.2012.09.001} {\bibfield
  {journal} {\bibinfo  {journal} {Prog. Part. Nucl. Phys.}\ }\textbf {\bibinfo
  {volume} {68}},\ \bibinfo {pages} {158} (\bibinfo {year} {2013})},\ \Eprint
  {http://arxiv.org/abs/1204.4617} {arXiv:1204.4617 [nucl-th]} \BibitemShut
  {NoStop}%
\bibitem [{\citenamefont {Ekstr{\"o}m}\ \emph {et~al.}(2023)\citenamefont
  {Ekstr{\"o}m}, \citenamefont {Forss{\'e}n}, \citenamefont {Hagen},
  \citenamefont {Jansen}, \citenamefont {Jiang},\ and\ \citenamefont
  {Papenbrock}}]{Ekstrom:2022yea}%
  \BibitemOpen
  \bibfield  {author} {\bibinfo {author} {\bibfnamefont {A.}~\bibnamefont
  {Ekstr{\"o}m}}, \bibinfo {author} {\bibfnamefont {C.}~\bibnamefont
  {Forss{\'e}n}}, \bibinfo {author} {\bibfnamefont {G.}~\bibnamefont {Hagen}},
  \bibinfo {author} {\bibfnamefont {G.~R.}\ \bibnamefont {Jansen}}, \bibinfo
  {author} {\bibfnamefont {W.}~\bibnamefont {Jiang}}, \ and\ \bibinfo {author}
  {\bibfnamefont {T.}~\bibnamefont {Papenbrock}},\ }\href {\doibase
  10.3389/fphy.2023.1129094} {\bibfield  {journal} {\bibinfo  {journal} {Front.
  Phys.}\ }\textbf {\bibinfo {volume} {11}},\ \bibinfo {pages} {1129094}
  (\bibinfo {year} {2023})},\ \Eprint {http://arxiv.org/abs/2212.11064}
  {arXiv:2212.11064 [nucl-th]} \BibitemShut {NoStop}%
\bibitem [{\citenamefont {Papenbrock}(2024)}]{Papenbrock:2024vhr}%
  \BibitemOpen
  \bibfield  {author} {\bibinfo {author} {\bibfnamefont {T.}~\bibnamefont
  {Papenbrock}}\ }(\bibinfo {year} {2024})\ \Eprint
  {http://arxiv.org/abs/2410.00843} {arXiv:2410.00843 [nucl-th]} \BibitemShut
  {NoStop}%
\bibitem [{\citenamefont {Wirth}\ \emph {et~al.}(2014)\citenamefont {Wirth},
  \citenamefont {Gazda}, \citenamefont {Navr{\'a}til}, \citenamefont {Calci},
  \citenamefont {Langhammer},\ and\ \citenamefont {Roth}}]{Wirth:2014apa}%
  \BibitemOpen
  \bibfield  {author} {\bibinfo {author} {\bibfnamefont {R.}~\bibnamefont
  {Wirth}}, \bibinfo {author} {\bibfnamefont {D.}~\bibnamefont {Gazda}},
  \bibinfo {author} {\bibfnamefont {P.}~\bibnamefont {Navr{\'a}til}}, \bibinfo
  {author} {\bibfnamefont {A.}~\bibnamefont {Calci}}, \bibinfo {author}
  {\bibfnamefont {J.}~\bibnamefont {Langhammer}}, \ and\ \bibinfo {author}
  {\bibfnamefont {R.}~\bibnamefont {Roth}},\ }\href {\doibase
  10.1103/PhysRevLett.113.192502} {\bibfield  {journal} {\bibinfo  {journal}
  {Phys. Rev. Lett.}\ }\textbf {\bibinfo {volume} {113}},\ \bibinfo {pages}
  {192502} (\bibinfo {year} {2014})},\ \Eprint {http://arxiv.org/abs/1403.3067}
  {arXiv:1403.3067 [nucl-th]} \BibitemShut {NoStop}%
\bibitem [{\citenamefont {Wirth}\ and\ \citenamefont
  {Roth}(2018)}]{Wirth:2017lso}%
  \BibitemOpen
  \bibfield  {author} {\bibinfo {author} {\bibfnamefont {R.}~\bibnamefont
  {Wirth}}\ and\ \bibinfo {author} {\bibfnamefont {R.}~\bibnamefont {Roth}},\
  }\href {\doibase 10.1016/j.physletb.2018.02.021} {\bibfield  {journal}
  {\bibinfo  {journal} {Phys. Lett. B}\ }\textbf {\bibinfo {volume} {779}},\
  \bibinfo {pages} {336} (\bibinfo {year} {2018})},\ \Eprint
  {http://arxiv.org/abs/1710.04880} {arXiv:1710.04880 [nucl-th]} \BibitemShut
  {NoStop}%
\bibitem [{\citenamefont {Le}\ \emph {et~al.}(2020)\citenamefont {Le},
  \citenamefont {Haidenbauer}, \citenamefont {Mei{\ss}ner},\ and\ \citenamefont
  {Nogga}}]{Le:2020zdu}%
  \BibitemOpen
  \bibfield  {author} {\bibinfo {author} {\bibfnamefont {H.}~\bibnamefont
  {Le}}, \bibinfo {author} {\bibfnamefont {J.}~\bibnamefont {Haidenbauer}},
  \bibinfo {author} {\bibfnamefont {U.-G.}\ \bibnamefont {Mei{\ss}ner}}, \ and\
  \bibinfo {author} {\bibfnamefont {A.}~\bibnamefont {Nogga}},\ }\href
  {\doibase 10.1140/epja/s10050-020-00314-6} {\bibfield  {journal} {\bibinfo
  {journal} {Eur. Phys. J. A}\ }\textbf {\bibinfo {volume} {56}},\ \bibinfo
  {pages} {301} (\bibinfo {year} {2020})},\ \Eprint
  {http://arxiv.org/abs/2008.11565} {arXiv:2008.11565 [nucl-th]} \BibitemShut
  {NoStop}%
\bibitem [{\citenamefont {Le}\ \emph {et~al.}(2023)\citenamefont {Le},
  \citenamefont {Haidenbauer}, \citenamefont {Mei{\ss}ner},\ and\ \citenamefont
  {Nogga}}]{Le:2022ikc}%
  \BibitemOpen
  \bibfield  {author} {\bibinfo {author} {\bibfnamefont {H.}~\bibnamefont
  {Le}}, \bibinfo {author} {\bibfnamefont {J.}~\bibnamefont {Haidenbauer}},
  \bibinfo {author} {\bibfnamefont {U.-G.}\ \bibnamefont {Mei{\ss}ner}}, \ and\
  \bibinfo {author} {\bibfnamefont {A.}~\bibnamefont {Nogga}},\ }\href
  {\doibase 10.1103/PhysRevC.107.024002} {\bibfield  {journal} {\bibinfo
  {journal} {Phys. Rev. C}\ }\textbf {\bibinfo {volume} {107}},\ \bibinfo
  {pages} {024002} (\bibinfo {year} {2023})},\ \Eprint
  {http://arxiv.org/abs/2210.03387} {arXiv:2210.03387 [nucl-th]} \BibitemShut
  {NoStop}%
\bibitem [{\citenamefont {Le}\ \emph {et~al.}(2025)\citenamefont {Le},
  \citenamefont {Haidenbauer}, \citenamefont {Mei{\ss}ner},\ and\ \citenamefont
  {Nogga}}]{Le:2024rkd}%
  \BibitemOpen
  \bibfield  {author} {\bibinfo {author} {\bibfnamefont {H.}~\bibnamefont
  {Le}}, \bibinfo {author} {\bibfnamefont {J.}~\bibnamefont {Haidenbauer}},
  \bibinfo {author} {\bibfnamefont {U.-G.}\ \bibnamefont {Mei{\ss}ner}}, \ and\
  \bibinfo {author} {\bibfnamefont {A.}~\bibnamefont {Nogga}},\ }\href
  {\doibase 10.1103/PhysRevLett.134.072502} {\bibfield  {journal} {\bibinfo
  {journal} {Phys. Rev. Lett.}\ }\textbf {\bibinfo {volume} {134}},\ \bibinfo
  {pages} {072502} (\bibinfo {year} {2025})},\ \Eprint
  {http://arxiv.org/abs/2409.18577} {arXiv:2409.18577 [nucl-th]} \BibitemShut
  {NoStop}%
\bibitem [{\citenamefont {Barnea}\ \emph {et~al.}(2006)\citenamefont {Barnea},
  \citenamefont {Vijande},\ and\ \citenamefont {Valcarce}}]{Barnea:2006sd}%
  \BibitemOpen
  \bibfield  {author} {\bibinfo {author} {\bibfnamefont {N.}~\bibnamefont
  {Barnea}}, \bibinfo {author} {\bibfnamefont {J.}~\bibnamefont {Vijande}}, \
  and\ \bibinfo {author} {\bibfnamefont {A.}~\bibnamefont {Valcarce}},\ }\href
  {\doibase 10.1103/PhysRevD.73.054004} {\bibfield  {journal} {\bibinfo
  {journal} {Phys. Rev. D}\ }\textbf {\bibinfo {volume} {73}},\ \bibinfo
  {pages} {054004} (\bibinfo {year} {2006})},\ \Eprint
  {http://arxiv.org/abs/hep-ph/0604010} {arXiv:hep-ph/0604010} \BibitemShut
  {NoStop}%
\bibitem [{\citenamefont {Zhao}\ and\ \citenamefont
  {Shi}(2024{\natexlab{a}})}]{Zhao:2023qww}%
  \BibitemOpen
  \bibfield  {author} {\bibinfo {author} {\bibfnamefont {J.}~\bibnamefont
  {Zhao}}\ and\ \bibinfo {author} {\bibfnamefont {S.}~\bibnamefont {Shi}},\
  }\href {\doibase 10.1103/PhysRevC.109.024901} {\bibfield  {journal} {\bibinfo
   {journal} {Phys. Rev. C}\ }\textbf {\bibinfo {volume} {109}},\ \bibinfo
  {pages} {024901} (\bibinfo {year} {2024}{\natexlab{a}})},\ \Eprint
  {http://arxiv.org/abs/2311.04594} {arXiv:2311.04594 [nucl-th]} \BibitemShut
  {NoStop}%
\bibitem [{\citenamefont {Zhao}\ and\ \citenamefont
  {Zhuang}(2017)}]{Zhao:2017gpq}%
  \BibitemOpen
  \bibfield  {author} {\bibinfo {author} {\bibfnamefont {J.}~\bibnamefont
  {Zhao}}\ and\ \bibinfo {author} {\bibfnamefont {P.}~\bibnamefont {Zhuang}},\
  }\href {\doibase 10.1007/s00601-017-1255-9} {\bibfield  {journal} {\bibinfo
  {journal} {Few Body Syst.}\ }\textbf {\bibinfo {volume} {58}},\ \bibinfo
  {pages} {100} (\bibinfo {year} {2017})}\BibitemShut {NoStop}%
\bibitem [{\citenamefont {Raynal}\ and\ \citenamefont
  {Revai}(1970)}]{Raynal:1970ah}%
  \BibitemOpen
  \bibfield  {author} {\bibinfo {author} {\bibfnamefont {J.}~\bibnamefont
  {Raynal}}\ and\ \bibinfo {author} {\bibfnamefont {J.}~\bibnamefont {Revai}},\
  }\href {\doibase 10.1007/BF02756127} {\bibfield  {journal} {\bibinfo
  {journal} {Nuovo Cim. A}\ }\textbf {\bibinfo {volume} {68}},\ \bibinfo
  {pages} {612} (\bibinfo {year} {1970})}\BibitemShut {NoStop}%
\bibitem [{\citenamefont {Zhao}\ and\ \citenamefont
  {Shi}(2024{\natexlab{b}})}]{Zhao:2024atb}%
  \BibitemOpen
  \bibfield  {author} {\bibinfo {author} {\bibfnamefont {J.}~\bibnamefont
  {Zhao}}\ and\ \bibinfo {author} {\bibfnamefont {S.}~\bibnamefont {Shi}},\
  }\href {\doibase 10.1016/j.cpc.2024.109284} {\bibfield  {journal} {\bibinfo
  {journal} {Comput. Phys. Commun.}\ }\textbf {\bibinfo {volume} {303}},\
  \bibinfo {pages} {109284} (\bibinfo {year} {2024}{\natexlab{b}})},\ \Eprint
  {http://arxiv.org/abs/2403.02747} {arXiv:2403.02747 [physics.comp-ph]}
  \BibitemShut {NoStop}%
\bibitem [{\citenamefont {Navas}\ \emph {et~al.}(2024)\citenamefont {Navas}
  \emph {et~al.}}]{ParticleDataGroup:2024cfk}%
  \BibitemOpen
  \bibfield  {author} {\bibinfo {author} {\bibfnamefont {S.}~\bibnamefont
  {Navas}} \emph {et~al.} (\bibinfo {collaboration} {Particle Data Group}),\
  }\href {\doibase 10.1103/PhysRevD.110.030001} {\bibfield  {journal} {\bibinfo
   {journal} {Phys. Rev. D}\ }\textbf {\bibinfo {volume} {110}},\ \bibinfo
  {pages} {030001} (\bibinfo {year} {2024})}\BibitemShut {NoStop}%
\bibitem [{\citenamefont {Lamia}\ \emph {et~al.}(2012)\citenamefont {Lamia},
  \citenamefont {La~Cognata}, \citenamefont {Spitaleri}, \citenamefont
  {Irgaziev},\ and\ \citenamefont {Pizzone}}]{Lamia:2012zz}%
  \BibitemOpen
  \bibfield  {author} {\bibinfo {author} {\bibfnamefont {L.}~\bibnamefont
  {Lamia}}, \bibinfo {author} {\bibfnamefont {M.}~\bibnamefont {La~Cognata}},
  \bibinfo {author} {\bibfnamefont {C.}~\bibnamefont {Spitaleri}}, \bibinfo
  {author} {\bibfnamefont {B.}~\bibnamefont {Irgaziev}}, \ and\ \bibinfo
  {author} {\bibfnamefont {R.~G.}\ \bibnamefont {Pizzone}},\ }\href {\doibase
  10.1103/PhysRevC.85.025805} {\bibfield  {journal} {\bibinfo  {journal} {Phys.
  Rev. C}\ }\textbf {\bibinfo {volume} {85}},\ \bibinfo {pages} {025805}
  (\bibinfo {year} {2012})}\BibitemShut {NoStop}%
\bibitem [{\citenamefont {Yukawa}(1935)}]{Yukawa:1935xg}%
  \BibitemOpen
  \bibfield  {author} {\bibinfo {author} {\bibfnamefont {H.}~\bibnamefont
  {Yukawa}},\ }\href {\doibase 10.1143/PTPS.1.1} {\bibfield  {journal}
  {\bibinfo  {journal} {Proc. Phys. Math. Soc. Jap.}\ }\textbf {\bibinfo
  {volume} {17}},\ \bibinfo {pages} {48} (\bibinfo {year} {1935})}\BibitemShut
  {NoStop}%
\bibitem [{\citenamefont {Bryan}\ and\ \citenamefont
  {Scott}(1964)}]{Bryan:1964zzb}%
  \BibitemOpen
  \bibfield  {author} {\bibinfo {author} {\bibfnamefont {R.~A.}\ \bibnamefont
  {Bryan}}\ and\ \bibinfo {author} {\bibfnamefont {B.~L.}\ \bibnamefont
  {Scott}},\ }\href {\doibase 10.1103/PhysRev.135.B434} {\bibfield  {journal}
  {\bibinfo  {journal} {Phys. Rev.}\ }\textbf {\bibinfo {volume} {135}},\
  \bibinfo {pages} {B434} (\bibinfo {year} {1964})}\BibitemShut {NoStop}%
\bibitem [{\citenamefont {Lacombe}\ \emph {et~al.}(1980)\citenamefont
  {Lacombe}, \citenamefont {Loiseau}, \citenamefont {Richard}, \citenamefont
  {Vinh~Mau}, \citenamefont {Cote}, \citenamefont {Pires},\ and\ \citenamefont
  {De~Tourreil}}]{Lacombe:1980dr}%
  \BibitemOpen
  \bibfield  {author} {\bibinfo {author} {\bibfnamefont {M.}~\bibnamefont
  {Lacombe}}, \bibinfo {author} {\bibfnamefont {B.}~\bibnamefont {Loiseau}},
  \bibinfo {author} {\bibfnamefont {J.~M.}\ \bibnamefont {Richard}}, \bibinfo
  {author} {\bibfnamefont {R.}~\bibnamefont {Vinh~Mau}}, \bibinfo {author}
  {\bibfnamefont {J.}~\bibnamefont {Cote}}, \bibinfo {author} {\bibfnamefont
  {P.}~\bibnamefont {Pires}}, \ and\ \bibinfo {author} {\bibfnamefont
  {R.}~\bibnamefont {De~Tourreil}},\ }\href {\doibase 10.1103/PhysRevC.21.861}
  {\bibfield  {journal} {\bibinfo  {journal} {Phys. Rev. C}\ }\textbf {\bibinfo
  {volume} {21}},\ \bibinfo {pages} {861} (\bibinfo {year} {1980})}\BibitemShut
  {NoStop}%
\bibitem [{\citenamefont {Wiringa}\ \emph {et~al.}(1995)\citenamefont
  {Wiringa}, \citenamefont {Stoks},\ and\ \citenamefont
  {Schiavilla}}]{Wiringa:1994wb}%
  \BibitemOpen
  \bibfield  {author} {\bibinfo {author} {\bibfnamefont {R.~B.}\ \bibnamefont
  {Wiringa}}, \bibinfo {author} {\bibfnamefont {V.~G.~J.}\ \bibnamefont
  {Stoks}}, \ and\ \bibinfo {author} {\bibfnamefont {R.}~\bibnamefont
  {Schiavilla}},\ }\href {\doibase 10.1103/PhysRevC.51.38} {\bibfield
  {journal} {\bibinfo  {journal} {Phys. Rev. C}\ }\textbf {\bibinfo {volume}
  {51}},\ \bibinfo {pages} {38} (\bibinfo {year} {1995})},\ \Eprint
  {http://arxiv.org/abs/nucl-th/9408016} {arXiv:nucl-th/9408016} \BibitemShut
  {NoStop}%
\bibitem [{\citenamefont {Reid}(1968)}]{Reid:1968sq}%
  \BibitemOpen
  \bibfield  {author} {\bibinfo {author} {\bibfnamefont {R.~V.}\ \bibnamefont
  {Reid}, \bibfnamefont {Jr.}},\ }\href {\doibase 10.1016/0003-4916(68)90126-7}
  {\bibfield  {journal} {\bibinfo  {journal} {Annals Phys.}\ }\textbf {\bibinfo
  {volume} {50}},\ \bibinfo {pages} {411} (\bibinfo {year} {1968})}\BibitemShut
  {NoStop}%
\bibitem [{\citenamefont {Stoks}\ \emph {et~al.}(1994)\citenamefont {Stoks},
  \citenamefont {Klomp}, \citenamefont {Terheggen},\ and\ \citenamefont
  {de~Swart}}]{Stoks:1994wp}%
  \BibitemOpen
  \bibfield  {author} {\bibinfo {author} {\bibfnamefont {V.~G.~J.}\
  \bibnamefont {Stoks}}, \bibinfo {author} {\bibfnamefont {R.~A.~M.}\
  \bibnamefont {Klomp}}, \bibinfo {author} {\bibfnamefont {C.~P.~F.}\
  \bibnamefont {Terheggen}}, \ and\ \bibinfo {author} {\bibfnamefont {J.~J.}\
  \bibnamefont {de~Swart}},\ }\href {\doibase 10.1103/PhysRevC.49.2950}
  {\bibfield  {journal} {\bibinfo  {journal} {Phys. Rev. C}\ }\textbf {\bibinfo
  {volume} {49}},\ \bibinfo {pages} {2950} (\bibinfo {year} {1994})},\ \Eprint
  {http://arxiv.org/abs/nucl-th/9406039} {arXiv:nucl-th/9406039} \BibitemShut
  {NoStop}%
\bibitem [{\citenamefont {Nagels}\ \emph {et~al.}(2019)\citenamefont {Nagels},
  \citenamefont {Rijken},\ and\ \citenamefont {Yamamoto}}]{Nagels:2014qqa}%
  \BibitemOpen
  \bibfield  {author} {\bibinfo {author} {\bibfnamefont {M.~M.}\ \bibnamefont
  {Nagels}}, \bibinfo {author} {\bibfnamefont {T.~A.}\ \bibnamefont {Rijken}},
  \ and\ \bibinfo {author} {\bibfnamefont {Y.}~\bibnamefont {Yamamoto}},\
  }\href {\doibase 10.1103/PhysRevC.99.044002} {\bibfield  {journal} {\bibinfo
  {journal} {Phys. Rev. C}\ }\textbf {\bibinfo {volume} {99}},\ \bibinfo
  {pages} {044002} (\bibinfo {year} {2019})},\ \Eprint
  {http://arxiv.org/abs/1408.4825} {arXiv:1408.4825 [nucl-th]} \BibitemShut
  {NoStop}%
\bibitem [{\citenamefont {Bodmer}\ \emph {et~al.}(1984)\citenamefont {Bodmer},
  \citenamefont {Usmani},\ and\ \citenamefont {Carlson}}]{Bodmer:1984gc}%
  \BibitemOpen
  \bibfield  {author} {\bibinfo {author} {\bibfnamefont {A.~R.}\ \bibnamefont
  {Bodmer}}, \bibinfo {author} {\bibfnamefont {Q.~N.}\ \bibnamefont {Usmani}},
  \ and\ \bibinfo {author} {\bibfnamefont {J.}~\bibnamefont {Carlson}},\ }\href
  {\doibase 10.1103/PhysRevC.29.684} {\bibfield  {journal} {\bibinfo  {journal}
  {Phys. Rev. C}\ }\textbf {\bibinfo {volume} {29}},\ \bibinfo {pages} {684}
  (\bibinfo {year} {1984})}\BibitemShut {NoStop}%
\bibitem [{\citenamefont {Otsuki}(1968)}]{10.1143/PTPS.42.39}%
  \BibitemOpen
  \bibfield  {author} {\bibinfo {author} {\bibfnamefont {S.}~\bibnamefont
  {Otsuki}},\ }\href {\doibase 10.1143/PTPS.42.39} {\bibfield  {journal}
  {\bibinfo  {journal} {Progress of Theoretical Physics Supplement}\ }\textbf
  {\bibinfo {volume} {42}},\ \bibinfo {pages} {39} (\bibinfo {year} {1968})},\
  \Eprint
  {http://arxiv.org/abs/https://academic.oup.com/ptps/article-pdf/doi/10.1143/PTPS.42.39/5346731/42-39.pdf}
  {https://academic.oup.com/ptps/article-pdf/doi/10.1143/PTPS.42.39/5346731/42-39.pdf}
  \BibitemShut {NoStop}%
\bibitem [{\citenamefont {McGee}(1967)}]{McGee:1967zza}%
  \BibitemOpen
  \bibfield  {author} {\bibinfo {author} {\bibfnamefont {I.~J.}\ \bibnamefont
  {McGee}},\ }\href {\doibase 10.1103/PhysRev.158.1500} {\bibfield  {journal}
  {\bibinfo  {journal} {Phys. Rev.}\ }\textbf {\bibinfo {volume} {158}},\
  \bibinfo {pages} {1500} (\bibinfo {year} {1967})}\BibitemShut {NoStop}%
\bibitem [{\citenamefont {Armstrong}\ \emph {et~al.}(2004)\citenamefont
  {Armstrong} \emph {et~al.}}]{E864:2002xhb}%
  \BibitemOpen
  \bibfield  {author} {\bibinfo {author} {\bibfnamefont {T.~A.}\ \bibnamefont
  {Armstrong}} \emph {et~al.} (\bibinfo {collaboration} {E864}),\ }\href
  {\doibase 10.1103/PhysRevC.70.024902} {\bibfield  {journal} {\bibinfo
  {journal} {Phys. Rev. C}\ }\textbf {\bibinfo {volume} {70}},\ \bibinfo
  {pages} {024902} (\bibinfo {year} {2004})},\ \Eprint
  {http://arxiv.org/abs/nucl-ex/0211010} {arXiv:nucl-ex/0211010} \BibitemShut
  {NoStop}%
\bibitem [{\citenamefont {Nemura}\ \emph {et~al.}(2000)\citenamefont {Nemura},
  \citenamefont {Suzuki}, \citenamefont {Fujiwara},\ and\ \citenamefont
  {Nakamoto}}]{Nemura:1999qp}%
  \BibitemOpen
  \bibfield  {author} {\bibinfo {author} {\bibfnamefont {H.}~\bibnamefont
  {Nemura}}, \bibinfo {author} {\bibfnamefont {Y.}~\bibnamefont {Suzuki}},
  \bibinfo {author} {\bibfnamefont {Y.}~\bibnamefont {Fujiwara}}, \ and\
  \bibinfo {author} {\bibfnamefont {C.}~\bibnamefont {Nakamoto}},\ }\href
  {\doibase 10.1143/PTP.103.929} {\bibfield  {journal} {\bibinfo  {journal}
  {Prog. Theor. Phys.}\ }\textbf {\bibinfo {volume} {103}},\ \bibinfo {pages}
  {929} (\bibinfo {year} {2000})},\ \Eprint
  {http://arxiv.org/abs/nucl-th/9912065} {arXiv:nucl-th/9912065} \BibitemShut
  {NoStop}%
\bibitem [{\citenamefont {Zhao}\ \emph {et~al.}(2024)\citenamefont {Zhao},
  \citenamefont {Gossiaux}, \citenamefont {Song}, \citenamefont
  {Bratkovskaya},\ and\ \citenamefont {Aichelin}}]{Zhao:2023dvk}%
  \BibitemOpen
  \bibfield  {author} {\bibinfo {author} {\bibfnamefont {J.}~\bibnamefont
  {Zhao}}, \bibinfo {author} {\bibfnamefont {P.~B.}\ \bibnamefont {Gossiaux}},
  \bibinfo {author} {\bibfnamefont {T.}~\bibnamefont {Song}}, \bibinfo {author}
  {\bibfnamefont {E.}~\bibnamefont {Bratkovskaya}}, \ and\ \bibinfo {author}
  {\bibfnamefont {J.}~\bibnamefont {Aichelin}},\ }\href {\doibase
  10.1051/epjconf/202429609014} {\bibfield  {journal} {\bibinfo  {journal} {EPJ
  Web Conf.}\ }\textbf {\bibinfo {volume} {296}},\ \bibinfo {pages} {09014}
  (\bibinfo {year} {2024})},\ \Eprint {http://arxiv.org/abs/2312.11349}
  {arXiv:2312.11349 [hep-ph]} \BibitemShut {NoStop}%
\bibitem [{\citenamefont {Chen}\ \emph {et~al.}(2003)\citenamefont {Chen},
  \citenamefont {Ko},\ and\ \citenamefont {Li}}]{Chen:2003ava}%
  \BibitemOpen
  \bibfield  {author} {\bibinfo {author} {\bibfnamefont {L.-W.}\ \bibnamefont
  {Chen}}, \bibinfo {author} {\bibfnamefont {C.~M.}\ \bibnamefont {Ko}}, \ and\
  \bibinfo {author} {\bibfnamefont {B.-A.}\ \bibnamefont {Li}},\ }\href
  {\doibase 10.1016/j.nuclphysa.2003.09.010} {\bibfield  {journal} {\bibinfo
  {journal} {Nucl. Phys. A}\ }\textbf {\bibinfo {volume} {729}},\ \bibinfo
  {pages} {809} (\bibinfo {year} {2003})},\ \Eprint
  {http://arxiv.org/abs/nucl-th/0306032} {arXiv:nucl-th/0306032} \BibitemShut
  {NoStop}%
\bibitem [{\citenamefont {Sun}\ and\ \citenamefont {Chen}(2015)}]{Sun:2015jta}%
  \BibitemOpen
  \bibfield  {author} {\bibinfo {author} {\bibfnamefont {K.-J.}\ \bibnamefont
  {Sun}}\ and\ \bibinfo {author} {\bibfnamefont {L.-W.}\ \bibnamefont {Chen}},\
  }\href {\doibase 10.1016/j.physletb.2015.10.056} {\bibfield  {journal}
  {\bibinfo  {journal} {Phys. Lett. B}\ }\textbf {\bibinfo {volume} {751}},\
  \bibinfo {pages} {272} (\bibinfo {year} {2015})},\ \Eprint
  {http://arxiv.org/abs/1509.05302} {arXiv:1509.05302 [nucl-th]} \BibitemShut
  {NoStop}%
\bibitem [{\citenamefont {Liu}\ \emph {et~al.}(2024{\natexlab{b}})\citenamefont
  {Liu}, \citenamefont {Hu}, \citenamefont {He}, \citenamefont {Shi},\ and\
  \citenamefont {Xie}}]{Liu:2024ilw}%
  \BibitemOpen
  \bibfield  {author} {\bibinfo {author} {\bibfnamefont {L.~K.}\ \bibnamefont
  {Liu}}, \bibinfo {author} {\bibfnamefont {C.~L.}\ \bibnamefont {Hu}},
  \bibinfo {author} {\bibfnamefont {X.~H.}\ \bibnamefont {He}}, \bibinfo
  {author} {\bibfnamefont {S.~S.}\ \bibnamefont {Shi}}, \ and\ \bibinfo
  {author} {\bibfnamefont {G.~N.}\ \bibnamefont {Xie}},\ }\href {\doibase
  10.1016/j.physletb.2024.138853} {\bibfield  {journal} {\bibinfo  {journal}
  {Phys. Lett. B}\ }\textbf {\bibinfo {volume} {855}},\ \bibinfo {pages}
  {138853} (\bibinfo {year} {2024}{\natexlab{b}})},\ \Eprint
  {http://arxiv.org/abs/2404.13582} {arXiv:2404.13582 [nucl-th]} \BibitemShut
  {NoStop}%
\end{thebibliography}%

\end{document}